\documentclass[12pt]{emulateapj}

\newcommand{\jo}{SDSS J0242+0049}
\newcommand{\kms}{km\,s$^{-1}$}
\newcommand{\cmss}{cm~s$^{-2}$}
\slugcomment{}
\journalinfo{Submitted to ApJ Dec. 15, 2006; Accepted April 27, 2007}
\shorttitle{Quasar Outflow Constraints}
\shortauthors{Hall et al.}

\begin{document}
\title{Acceleration and Substructure Constraints in a Quasar Outflow}
\author{
Patrick B. Hall,\altaffilmark{1}
Sarah I. Sadavoy,\altaffilmark{1}
Damien Hutsemekers,\altaffilmark{2}
John E. Everett,\altaffilmark{3}
Alireza Rafiee\altaffilmark{1}
}
\altaffiltext{1}{Department of Physics and Astronomy,
York University, 4700 Keele St., Toronto, Ontario M3J 1P3, Canada}
\altaffiltext{2}{Senior Research Associate FNRS, University of Li\`ege,
     All\'ee du 6 ao\^ut 17, Bat. 5c, 4000 Li\`ege, Belgium}
\altaffiltext{3}{Departments of Physics and Astronomy and Center for Magnetic
Self-Organization in Laboratory and Astrophysical Plasmas, University of
Wisconsin-Madison, 1150 University Avenue, Madison, Wisconsin 53706, USA}

\begin{abstract}
We present observations of probable line-of-sight acceleration of a broad
absorption trough of \ion{C}{4} in the quasar SDSS J024221.87+004912.6.  
We also discuss how the velocity overlap of two other outflowing
systems in the same object constrains the properties of the outflows.
The \ion{Si}{4} doublet in each system has one unblended transition
and one transition which overlaps with absorption from the other system.
The residual flux in the overlapping trough
is well fit by the product of the residual fluxes in the unblended troughs.
For these optically thick systems to yield such a result, at least one of them
must consist of individual subunits rather than being a
single structure with velocity-dependent coverage of the source.
If these subunits are identical, opaque, spherical clouds, we
estimate the cloud radius to be $r\simeq 3.9\times 10^{15}$~cm.
If they are identical, opaque, linear filaments, we estimate their width to be
$w\simeq 6.5\times 10^{14}$~cm.  These subunits are observed to
cover the \ion{Mg}{2} broad emission line region of the quasar, at which
distance from the black hole the above filament width is equal
to the predicted scale height of the outer atmosphere of a thin accretion disk.
Insofar as that scale height is a natural size scale for structures originating
in an accretion disk, these observations are evidence that the accretion disk
can be a source of quasar absorption systems.
Based on data from ESO program 075.B-0190(A).
\end{abstract}

\section{Introduction}

Absorption systems in active galactic nuclei (AGN) can be classified as
\emph{intrinsic}, which are
associated with the active nucleus and
often outflowing from it, and \emph{intervening}, which originate from clouds
external to the AGN's environment.  Determining an absorption
system's classification can be difficult.
A reliable indication of intrinsic absorption is time variability,
such as a shift in the velocity of a given 
feature or changes in its absorption strength as a function of velocity.
In the broad absorption line (BAL) troughs $\gtrsim1000$\,km\,s$^{-1}$ wide
which are most often found in the luminous AGN known as quasars, 
reports of time variable absorption strengths have been relatively common
(\nocite{bea85,fol87,vsb87,sp88,tea88,bjb89,bjb92,bea92,bar94,hea95,mon96,hbj97,bhs97,hea97,sdss123,gea04,mect05}{Bromage} {et~al.} 1985; {Foltz} {et~al.} 1987; {Voit}, {Shull}, \& {Begelman} 1987; {Smith} \& {Penston} 1988; {Turnshek} {et~al.} 1988; {Barlow}, {Junkkarinen}, \&  {Burbidge} 1989, 1992a; {Barlow} {et~al.} 1992b; {Barlow} 1994; {Hamann} {et~al.} 1995; {Michalitsianos}, {Oliversen}, \&  {Nichols} 1996; {Hamann}, {Barlow}, \&  {Junkkarinen} 1997b; {Barlow}, {Hamann}, \& {Sargent} 1997; {Hamann} {et~al.} 1997a; {Hall} {et~al.} 2002; {Gallagher} {et~al.} 2004; {Misawa} {et~al.} 2005; Lundgren et al. 2006).
Such variability can even include the appearance or disappearance of absorption
systems \nocite{kea96,gce01,ma02,gea05,lchg05}({Koratkar} {et~al.} 1996; {Ganguly}, {Charlton}, \&  {Eracleous} 2001; {Ma} 2002; {Gallagher} {et~al.} 2005; {Leighly} {et~al.} 2005).  In contrast,
velocity shifts in BAL outflows have been reported in only
Q~1303+308 \nocite{vi01}({Vilkoviskij} \& {Irwin} 2001) and
Mrk~231 (\nocite{rvs02}{Rupke}, {Veilleux}, \& {Sanders} 2002, and references therein),
although \nocite{gea03}{Gabel} {et~al.} (2003) have observed deceleration of a narrow absorber in 
the Seyfert~1 NGC 3783.

Acceleration must occur for AGN outflows to reach their observed velocities. 
Nonetheless, velocity shifts in AGN outflows are seen quite rarely because 
acceleration of an AGN outflow does not automatically translate into a change 
in its observed velocity profile, and vice versa.  For example, a fixed 
mass loss rate into an outflow with a time-invariant driving force would yield 
a time-invariant acceleration profile with distance in the outflow,
and thus produce unchanging absorption troughs.
\nocite{aea99}{Arav} {et~al.} (1999) illustrate how radial acceleration of gas crossing our line of
sight with a non-negligible transverse velocity produces an observed absorption
trough with a broadened radial velocity profile that does not change with time.
Since our lines of sight to AGN are essentially radial, and since
AGN are fed by accretion disks consisting of gas with predominantly
orbital velocities, most AGN outflows are expected to have non-negligible
transverse as well as radial velocities.  Thus, most intrinsic absorbers
likely {\em are} exhibiting acceleration, disguised as a trough broader
than the thermal or turbulent velocity width of the gas.

What are we then to make of cases where an outflow {\em does} exhibit a velocity
shift?  First, note that when
our line of sight intersects the origin of an outflow, the
absorption trough can start at zero line-of-sight velocity in the AGN rest
frame, at least for ions present at the origin of the outflow.
Ions present only downstream in an outflow, or lines of sight intersecting an
outflow only downstream from its origin due to curvature in the flow lines,
will produce `detached' absorption troughs which do not start at zero velocity,
as will a shell of material ejected in an intermittent outflow.
With that in mind, consider possible explanations for
a velocity shift observed in a detached absorption trough.
Such a shift can be produced by changes in the ionization state as a function
of velocity in a fixed outflow, by changes in the acceleration profile
or geometry (or both) of such an outflow due to changes in the driving force or
mass loss rate, or by actual line-of-sight acceleration of a shell of material
from an intermittent outflow.  
Observations of velocity shifts are therefore worthwhile because they may yield
insights into specific scenarios for quasar absorbers.

Here we present multiple-epoch observations (\S 2) of a quasar in which a broad
absorption line trough of \ion{C}{4} increased in outflow velocity over 
1.4 rest-frame years (\S 3).  We also discuss how two overlapping
outflows in the same quasar provide constraints on the properties of those 
outflows (\S 4).  We end with our conclusions in \S 5.

\section{Observations}

The Sloan Digital Sky Survey \nocite{yor00}(SDSS; {York} {et~al.} 2000) is using a drift-scanning
camera \nocite{gun98}({Gunn} {et~al.} 1998) on a 2.5-m telescope \nocite{gun06}({Gunn} {et~al.} 2006) to image 10$^4$\,deg$^2$
of sky on the SDSS $ugriz$ AB magnitude system 
\nocite{fuk96,sdss82,sdss105,sdss153,ive04}({Fukugita} {et~al.} 1996; {Hogg} {et~al.} 2001; {Smith} {et~al.} 2002; {Pier} {et~al.} 2003; {Ivezi{\'c}} {et~al.} 2004).
Two multi-fiber, double spectrographs are being used to obtain resolution
$R\sim1850$ spectra covering $\simeq$3800-9200\,\AA\ 
for $\sim$10$^6$ galaxies to $r=17.8$ and $\sim$10$^5$ quasars to
$i=19.1$ ($i=20.2$ for $z>3$ candidates; \nocite{sdssqtarget}{Richards} {et~al.} 2002).

The $z_{em}=2.062$ BAL quasar SDSS J024221.87+004912.6 
\nocite{sdssedrq,sdssbalcat,dr3q,trump06}({Schneider} {et~al.} 2002; {Reichard} {et~al.} 2003; {Schneider} {et~al.} 2005; {Trump} {et~al.} 2006), hereafter referred to as \jo, 
was observed spectroscopically three times by the SDSS (Table \ref{spec}).
We selected it for high-resolution spectroscopic followup because of the
possible presence of narrow absorption in excited-state \ion{Si}{2} and
\ion{C}{2} at $z=2.042$.  A spectrum obtained with 
the ESO Very Large Telescope (VLT) Unit 2 (Kueyen)
and Ultra-Violet Echelle Spectrograph (UVES; \nocite{uves}{Dekker} {et~al.} 2000) 
confirms the presence of narrow, low-ionization absorption at that 
redshift,\footnote{The weak, narrow character of that absorption 
led to the classification of this object as a high-ionization BAL quasar
by \nocite{sdssbalcat}{Reichard} {et~al.} (2003) and \nocite{trump06}{Trump} {et~al.} (2006) based on its SDSS spectrum.}  
analysis of which will be reported elsewhere.

\begin{figure}
\epsscale{1.00}
\plotone{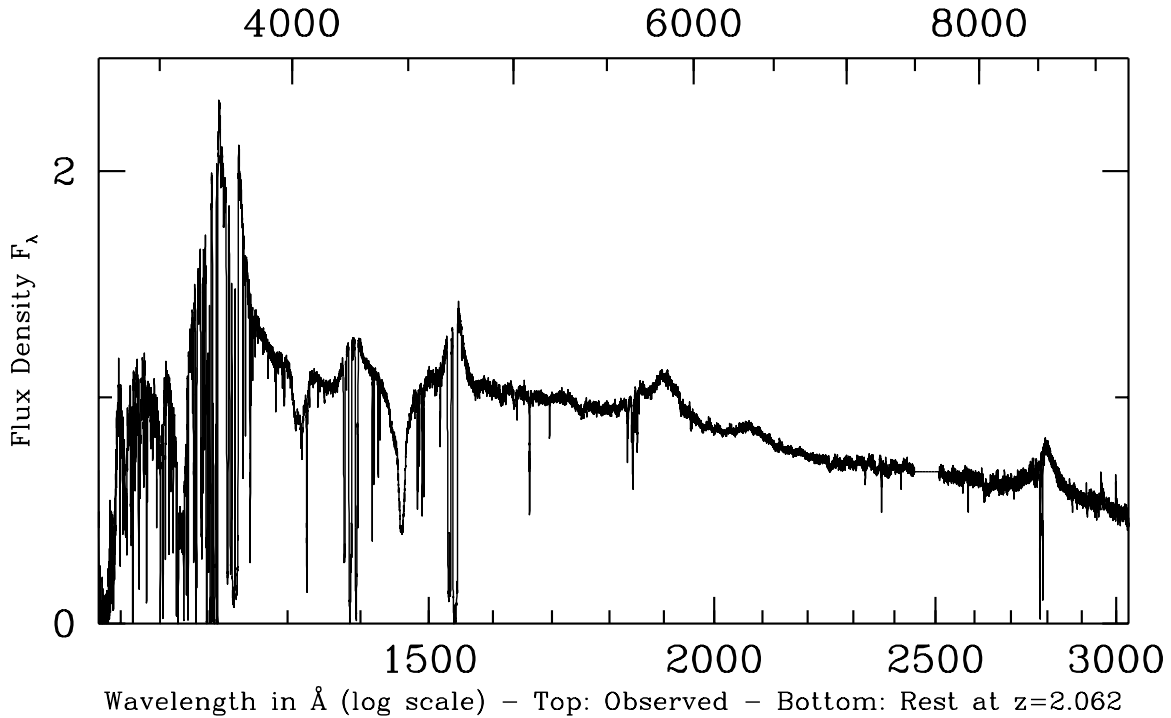}
\caption{VLT UT2 + UVES spectrum of \jo, smoothed by a 1\,\AA\ boxcar filter.}\label{figtot}
\end{figure}

We observed \jo\ with UVES on the VLT UT2 on the nights of 4-5 September 2005
through a 1\arcsec\ slit with 2x2 binning of the CCD, yielding $R\simeq40000$.
The weather ranged from clear to thin cirrus, with $0.8-1.0$\arcsec\ seeing.
\jo\ was observed for a total of 5.75 hours in two different spectral 
settings, yielding coverage from 3291-7521\,\AA\ and 7665-9300\,\AA.
Each exposure was reduced individually with optimum extraction \nocite{hor86}({Horne} 1986),
including simultaneous background and sky subtraction.  Telluric absorption
lines were removed for the red settings using observations of telluric standard
stars.  A weighted co-addition of the three exposures of each spectral setting
was performed with rejection of cosmic rays and known CCD artifacts.
Finally, all settings were rebinned to a vacuum heliocentric wavelength scale,
scaled in intensity by their overlap regions, and merged into a single spectrum
with a constant wavelength interval of 0.08\,\AA\ (Figure \ref{figtot}).
The SDSS spectra all share a common wavelength system with pixels equally
spaced in velocity, and so for ease of comparison we created a version of the
UVES spectrum binned to the those same wavelengths 
but not smoothed to the SDSS resolution.

\section{Broad Absorption Line Trough Velocity Shifts}\label{bal}

\begin{figure}
\plotone{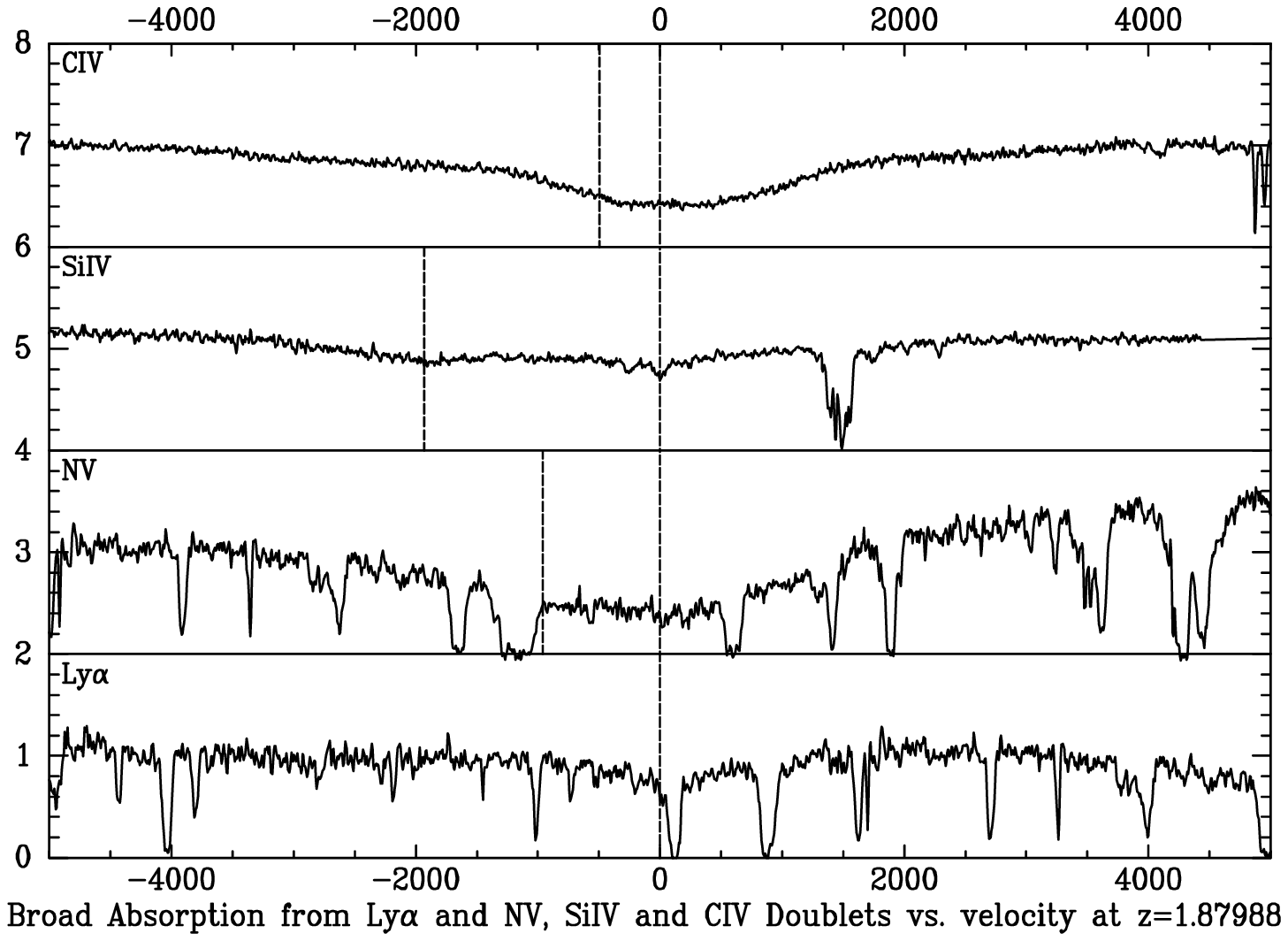}
\caption{UVES spectra of
BAL troughs in \jo\ vs. velocity (in \kms) in the $z=1.87988$ frame.
Negative velocities indicate blueshifts and positive velocities indicate
redshifts relative to that frame.
Zero velocity corresponds to the long-wavelength members of doublets,
and dashed vertical lines indicate all components of each transition.
Contaminating narrow absorption lines are present near all troughs,
but especially in those found shortward of the Ly$\alpha$ forest.}
\label{fig:BAL}
\end{figure}

The broadest absorption lines in \jo\ occur at a 
redshift $z\simeq1.87988$ ($v=-18400$\,\kms\ relative to the quasar)
in Ly$\alpha$, \ion{N}{5}, \ion{Si}{4} and \ion{C}{4} (Figure \ref{fig:BAL}).
There is an offset between the peak absorption in \ion{C}{4} and \ion{Si}{4}.
The redshift $z=1.87988$ was determined from the deepest absorption in the
\ion{Si}{4} trough, and does not match the deepest \ion{C}{4} absorption.
This can be ascribed to a changing ionization state in the outflow
as a function of velocity.

Comparison of the SDSS and UVES spectra suggested a shift in the position of
the \ion{C}{4} trough at this redshift.  To investigate further, continuum
regions around that trough and the \ion{Si}{4} trough at the same redshift
were fitted and used to normalize all observed spectra.
(The Ly$\alpha$ and \ion{N}{5} troughs lie outside the SDSS wavelength range.)
For each epoch, the \ion{C}{4} and \ion{Si}{4} regions were fit separately
with third order Legendre functions using
{\sc splot} in IRAF.\footnote{The Image Reduction and Analysis Facility (IRAF)
is distributed by the National Optical Astronomy Observatories, which is
operated by AURA, Inc., under contract to the National Science Foundation.}
The continuum sample windows were selected to avoid
emission lines in the quasar rest frame \nocite{sdss73}({Vanden Berk} {et~al.} 2001).  

The extent of any shift can be measured by minimizing the $\chi^{2}$ 
between the normalized pixel-by-pixel fluxes in the spectra 
when shifted by an integer number of pixels $m$
(assuming pixels equally spaced in velocity): 
\begin{equation}
\chi_{\nu,m}^2 = \frac{1}{N-m} \sum_{i=1}^{N-m} \frac{(f_{2,i}-f_{1,i+m})^2}{\sigma_{1,i}^2+\sigma_{2,i+m}^2} \label{eqChi}
\end{equation}
where $f_{2,i}$ and $f_{1,i+m}$ represent the flux in spectra from epochs 1 and
2 at pixels $i$ and $i+m$, respectively, $N$ is the
total number of pixels extracted from each spectrum for comparison
and $\sigma$ is the error for the flux at each pixel. 

\begin{figure}
\plotone{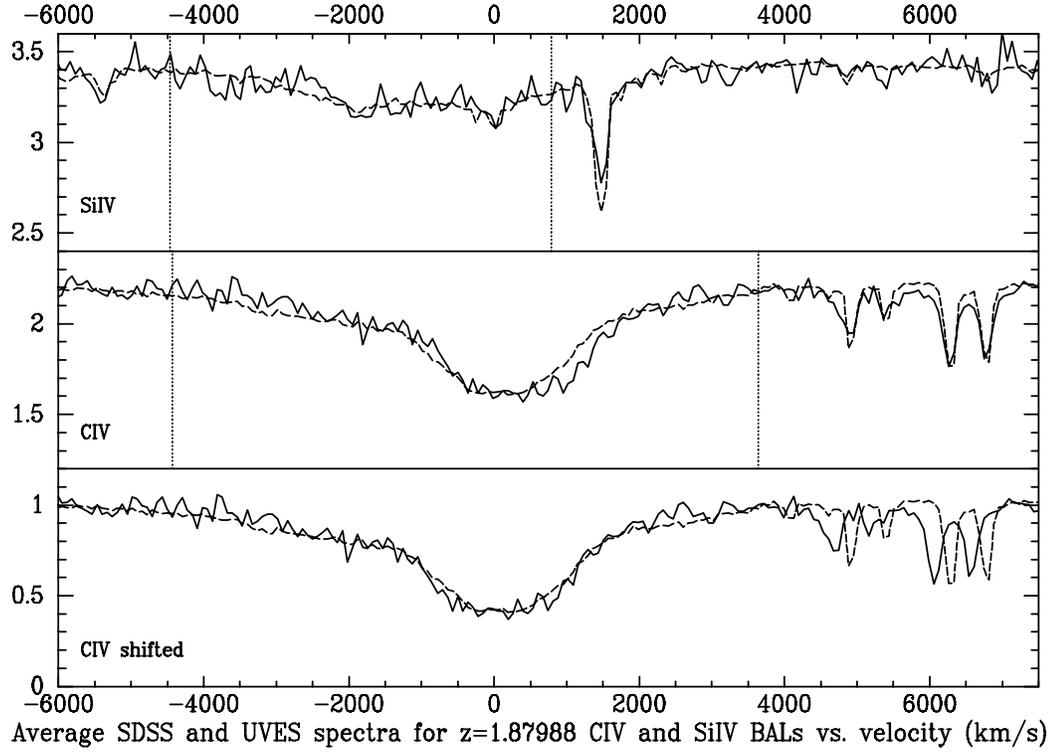}
\caption{Comparison of the $z=1.87988$ \ion{C}{4} BAL in \jo\ at the average
SDSS epoch and the UVES epoch.
Negative velocities indicate blueshifts and positive velocities redshifts,
relative to $z=1.87988$.
The solid line is a weighted average of all three SDSS spectra.  The dashed
line is the UVES spectrum binned into the same pixels as the SDSS spectra. 
Dotted vertical lines indicate the fitting regions used when conducting the
$\chi^{2}$ test. 
The top panel compares the unshifted spectra for the \ion{Si}{4} trough,
and the middle panel the unshifted spectra for the \ion{C}{4} trough.
The bottom panel compares the \ion{C}{4} troughs after shifting the average
SDSS spectrum toward shorter wavelengths by 3 pixels.}
\label{accNorm}
\end{figure}

The SDSS spectra from epochs 51821 and 
52188\footnote{Since the SDSS spectra from MJD 52177 and MJD 52199 are noisier
at short wavelengths than the SDSS spectrum from MJD 51821 and since visual 
inspection of them revealed no obvious difference in their BAL troughs, 
a weighted co-add of their spectra was made, with mean epoch 52188.}
were compared with the UVES spectrum from epoch 53619 (Table \ref{spec}).
A clear shift was found in \ion{C}{4} and a potentially
smaller shift in \ion{Si}{4}.
Neither trough shows a detectable shift between the SDSS spectra 
from epoch 51821 and epoch 52188, and neither would be expected to do so if
the observed long-term shift was due to a constant acceleration (the shift
between those two epochs would be $\lesssim 0.5$ pixel for \ion{C}{4}).
In light of this, the $\chi^2$ test was conducted again, using
a weighted average of all three SDSS spectra, with mean epoch 52066.
From that comparison we conclude that the shift in \ion{C}{4} is 
$3 \pm 1$ pixels
with 95.4\% confidence (2$\sigma$).
Zero velocity shift in \ion{C}{4} can be excluded with 99.9998\% confidence.
For \ion{Si}{4}, the shift is $1 \pm 3$ pixels at 95.4\% confidence.
Plots of these spectra are shown in the top two panels of Figure \ref{accNorm}.
It is important to note that there is no shift in the nearby narrow absorption
lines.  Also, both troughs appear to keep a relatively constant intensity, 
within the uncertainties.  The bottom panel of Figure \ref{accNorm} 
shows the excellent match to the epoch 53619 UVES spectrum
that results when the epoch 52066 average SDSS spectrum is shifted by 3 pixels.

The middle panel of Figure \ref{accNorm} may suggest
that the long-wavelength end of the \ion{C}{4}
trough has a greater shift than the short-wavelength end.  Splitting the
\ion{C}{4} trough into two sections, we find that $\chi^{2}$ is minimized at
a shift of $2^{+2}_{-1}$ pixels for the short-wavelength end and a shift of 
$4^{+1}_{-2}$ pixels for the long-wavelength edge,
but that a uniform shift produces a marginally lower minimum overall $\chi^{2}$.
Thus, while there is possible evidence for a nonuniform velocity
shift of the \ion{C}{4}
BAL trough, the current data are of insufficient quality to prove its existence.
Many physical effects could produce a nonuniform shift (expansion of an
overpressured, accelerated shell of gas from an intermittent outflow,
to give one example).

A shift of one SDSS pixel corresponds to a velocity shift of 69 \kms\ in the
observed frame or 22.5 \kms\ in the quasar rest frame ($z=2.062$).
A shift of $3 \pm 1$ SDSS pixels (2$\sigma$) over a rest-frame time span 
of 1.39 years thus gives an acceleration of 
$a = 0.154 \pm 0.025 \mbox{\ cm\ s}^{-2}$, where the error is 1$\sigma$.
Previously claimed accelerations for BAL troughs are much lower than that, at
$a = 0.035\pm 0.016$ \cmss\ over 5.5 rest-frame years in Q~1303+308 \nocite{vi01}({Vilkoviskij} \& {Irwin} 2001)
and $a=0.08 \pm 0.03$ \cmss\ over 12 rest-frame years for Mrk~231 \nocite{rvs02}({Rupke} {et~al.} 2002).
Our observation is more similar to that of \nocite{gea03}{Gabel} {et~al.} (2003), who determined the
deceleration of \ion{C}{4}, \ion{N}{5} and \ion{Si}{4} 
in a narrow absorption system in a Seyfert galaxy and found
(for \ion{C}{4}) relatively large values of $a=-0.25\pm 0.05$ \cmss\ and 
$a=-0.10\pm 0.03$ \cmss\ over 0.75 and 1.1 rest-frame years, respectively.
All of those observations involved much narrower troughs than is the case in
\jo.  Also, the 1$\sigma$ relative uncertainty associated with the acceleration
of \jo\ is lower than the previous BAL measurements.  These factors make \jo\ a
robust case for line-of-sight acceleration of a true BAL trough.
Still, it should be kept in mind that all these accelerations are
much smaller than the $a\simeq 100 \mbox{\ cm\ s}^{-2}$ predicted for 
the main acceleration phase of a disk wind in the model of \nocite{mur95}{Murray} {et~al.} (1995).

Furthermore, 
BAL troughs can vary for several reasons.  These include
acceleration or deceleration along the line of sight of some or all of the
absorbing gas,
a change in the ionization state of some or all of the gas, or
a change in $C(v)$ --- the covering factor of the gas as a function of the 
line-of-sight velocity ---
due to the movement of gas into or out of our line of sight, for example
due to a change in flow geometry (see the introduction and
\S 3.3 of \nocite{gea03}{Gabel} {et~al.} 2003).
In many cases of variability all of the above origins are possible, but there
are cases where acceleration is very unlikely to be the cause (see below).
Because of this,
to be conservative we cannot assume that BAL trough variability is due to 
acceleration 
even though acceleration {\em could} be the cause of much of the
observed variability.

Fig. 2 of \nocite{bjb89}{Barlow} {et~al.} (1989) and Fig. 2 of \nocite{bea92}{Barlow} {et~al.} (1992b) 
are cases where observed time variability of BAL troughs is almost certainly 
due to a change in the column densities of an ion at certain velocities 
(whether due to a changing ionization or to bulk motion into the line of sight),
not due to a given ionic column density changing its velocity.
More ambiguous cases are illustrated by
\ion{C}{4} in Q~1246$-$057 (Fig. 3 of \nocite{sp88}{Smith} \& {Penston} 1988)
and \ion{Si}{4} in Q~1413+117 (Fig. 15 of \nocite{tea88}{Turnshek} {et~al.} 1988).
In both of those cases, a second-epoch spectrum shows more absorption 
at the short-wavelength edge of the trough in question.
That could be because gas at lower outflow velocities in the trough was 
accelerated to higher velocities.  Yet in both cases, the
trough away from the short-wavelength edge is unchanged between the two epochs.
If acceleration was the cause of the variability, 
a reduction in covering factor or optical depth, or both, 
might be expected at the lower velocities where the gas originated.  
No reduction is seen, arguing against the
line-of-sight acceleration hypothesis for these cases of trough variability.

While every case for acceleration in a BAL trough will be ambiguous at some 
level, comparing the variability we report in \jo\ to previous cases leads
us to believe that ours is the least ambiguous case seen to date of
acceleration in a true BAL trough ($\gtrsim 1000$\ \kms\ wide).
Monitoring the future behavior of the $z=1.87988$
absorption in this quasar would be very worthwhile, to see if the acceleration
was temporary, is constant, increasing, or decreasing, or varies stochastically.
The latter might occur if the velocity shift is due to a variable flow geometry
or to ionization variations as a function of velocity
caused by a fluctuating ionizing luminosity.
(Recall from Figure 2 that this system shows some evidence for ionization
stratification with velocity, in the form of an offset between the
velocities of the peak \ion{Si}{4} and \ion{C}{4} absorption.)
As this quasar is located in the equatorial stripe of the SDSS, which has been
repeatedly imaged over the past 7 years, it should eventually be possible to
search for a correlation between its ultraviolet luminosity and the
acceleration of this system.
(From the spectra alone, there appears to be a 5-10\% increase in the 
luminosity of the object over the time spanned by the three SDSS spectra, 
but no information is available on longer timescales since 
the UVES spectrum is not spectrophotometrically calibrated.)
BAL trough velocity shifts are also expected if
BAL quasars are
a short-lived phase during which material is expelled from the nuclear region
\nocite{vwk93}({Voit}, {Weymann}, \& {Korista} 1993).  In such a model the accelerating trough in \jo\ could be 
interpreted as gas unusually close to the quasar,
currently experiencing an unusually large radiative acceleration.

\section{Overlapping \ion{Si}{4} Troughs}

There is a possible case of line-locking involving \ion{Si}{4}
in \jo.  Stable line-locking in a given doublet occurs 
when two conditions are met.  First,
the velocity separation between two absorption systems at different redshifts
must be
very nearly equal to the velocity separation of the two lines of a doublet
seen in both systems \nocite{bm89}({Braun} \& {Milgrom} 1989).  Second,
the reduction in line-driven acceleration of the shadowed system due to the
reduced incident flux in one component of the doublet must result in its 
acceleration being the same as that of the shadowing system.  This latter
condition may be difficult to meet in AGN outflows, where many lines
contribute to the radiative acceleration and there may also be substantial
non-radiative acceleration.  Nonetheless, some spectacular examples of
apparent line-locking in AGN do suggest that it can in fact occur 
(e.g., \nocite{splh02}{Srianand} {et~al.} 2002), even if only rarely.

\begin{figure}
\plotone{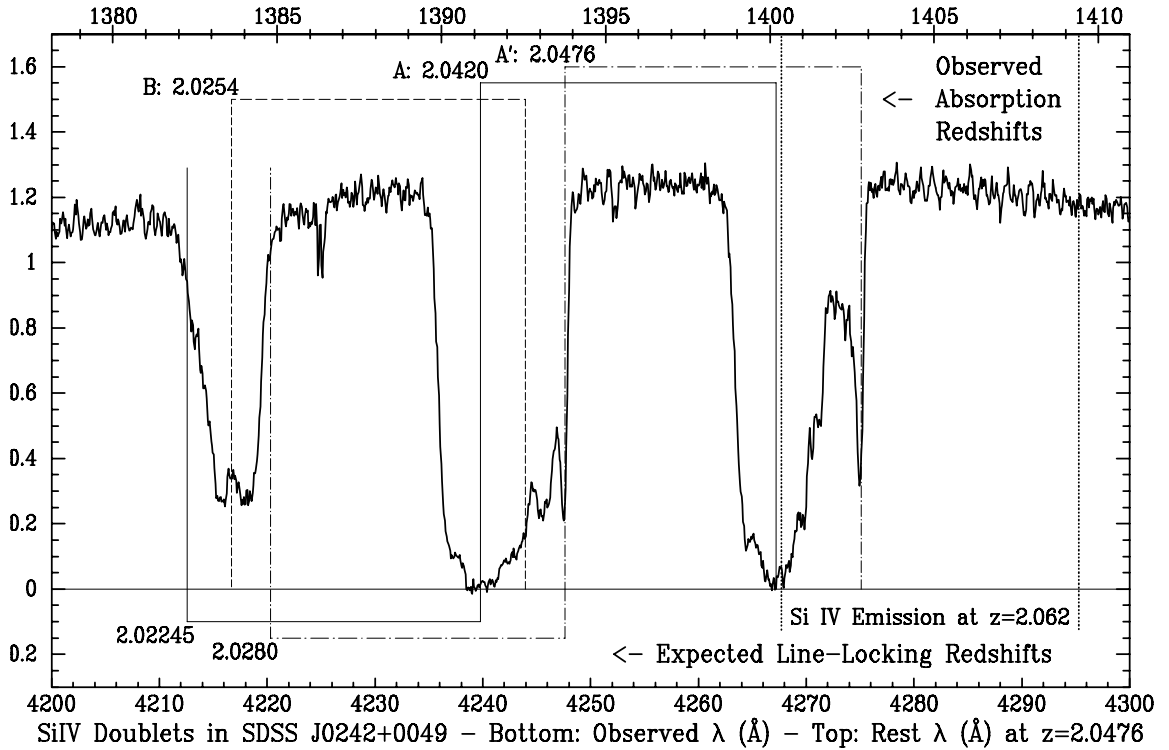}
\caption{Two broad, overlapping \ion{Si}{4} doublets in the unnormalized 
spectrum of \jo.  Line 
identifications and redshifts for the different troughs are given on the figure.
There is also narrow \ion{Si}{4} absorption at z=2.0314 which is not marked.
}\label{fig:lambda}
\end{figure}

As shown in Figure \ref{fig:lambda}, in \jo\ there is narrow \ion{Si}{4}
absorption at $z=2.0476$ (hereafter system A$'$)
and a broad \ion{Si}{4} trough centered at about $z=2.042$ (hereafter system A).
\ion{Si}{4} line-locking of a third absorption system to system A$'$ or A 
would result in absorption 1931 \kms\ shortward of those redshifts, 
at $z=2.0280$ or $z=2.02245$ respectively.
What is observed in the spectrum, however, is broad absorption in between the
expected redshifts, centered at $z=2.0254$ (hereafter system B).  
Both systems are observed in other transitions as well, 
with system B having more absorption
in \ion{N}{5} and \ion{C}{4} but less in \ion{S}{4} and \ion{Mg}{2}.

In this section we consider first the optical depths and covering factors of
these overlapping systems, with intriguing results.
We then consider whether they could be line-locked 
or in the process of becoming line-locked.

\subsection{\ion{Si}{4} Trough Optical Depths and Covering Factors}

\begin{figure}
\plotone{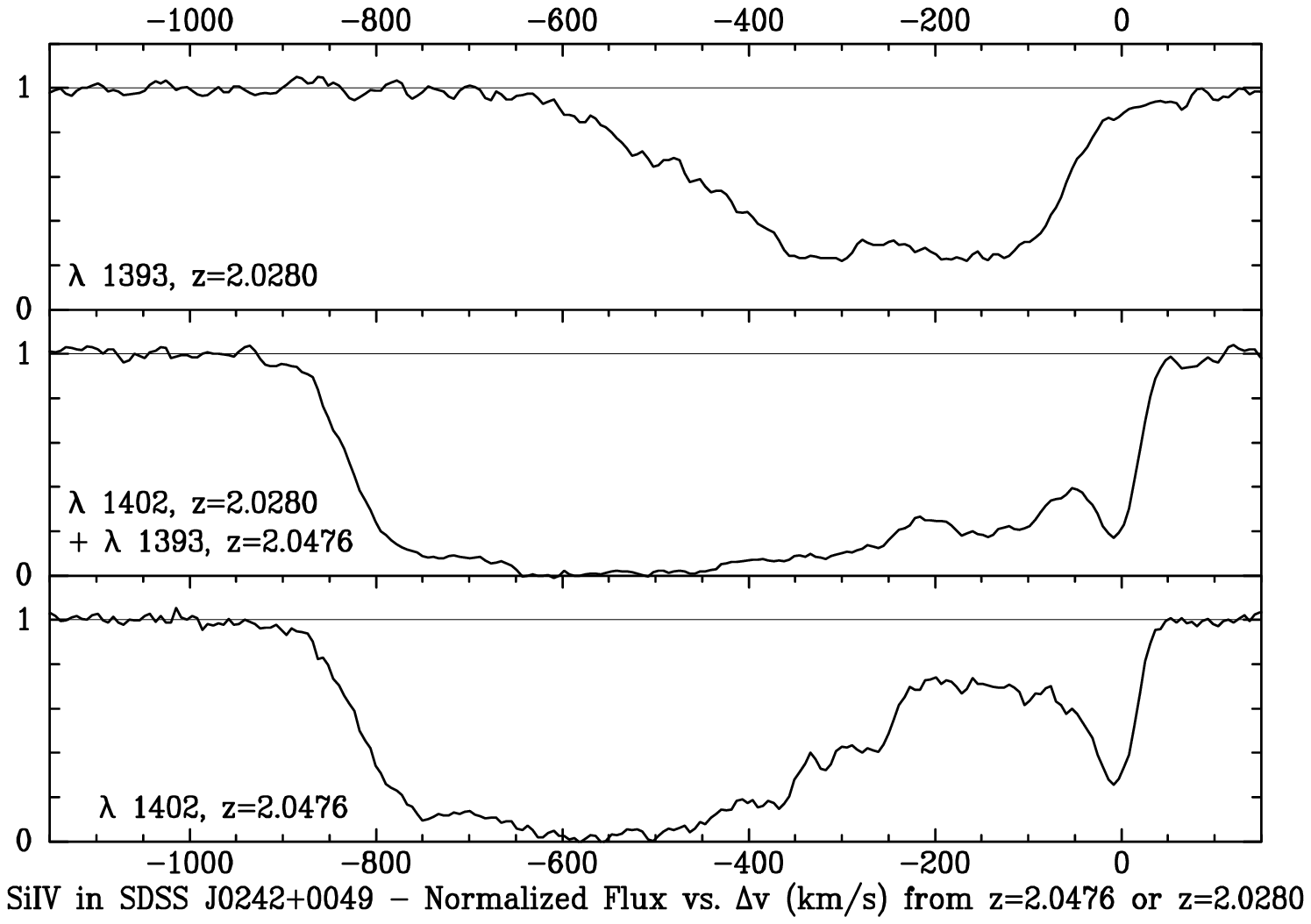}
\caption{Velocity plot of \ion{Si}{4} absorption after normalization by 
a fit to the total spectrum (continuum and weak emission lines).}
\label{fig:norm}
\end{figure}

It is useful to determine if the \ion{Si}{4} troughs under consideration
are optically thick or not.
Figure \ref{fig:norm} shows the absorption profiles in velocity space relative
to $z=2.0476$ or to the corresponding line-locked redshift of $z=2.0280$.
System A+A$'$, seen unblended in the bottom panel, is free from 
contamination in the blended trough (middle panel) at $-900<v<-650$ \kms.  
At those velocities, absorption from the $\lambda$1402 component of System
A+A$'$ (bottom panel) appears so similar in shape and intensity to absorption 
from the intrinsically stronger $\lambda$1393 component (middle panel)
that we can conclude system A+A$'$ is optically thick in \ion{Si}{4}.
For system B (seen unblended in the top panel) we
must
see how well various combinations of optical depth,
covering factor, and geometry \nocite{rvs05}({Rupke}, {Veilleux}, \& {Sanders} 2005)
can reproduce the profile of the trough composed of blended absorption from
system B and the optically thick system A+A$'$ (middle panel).

\begin{figure}
\epsscale{1.25}
\plotone{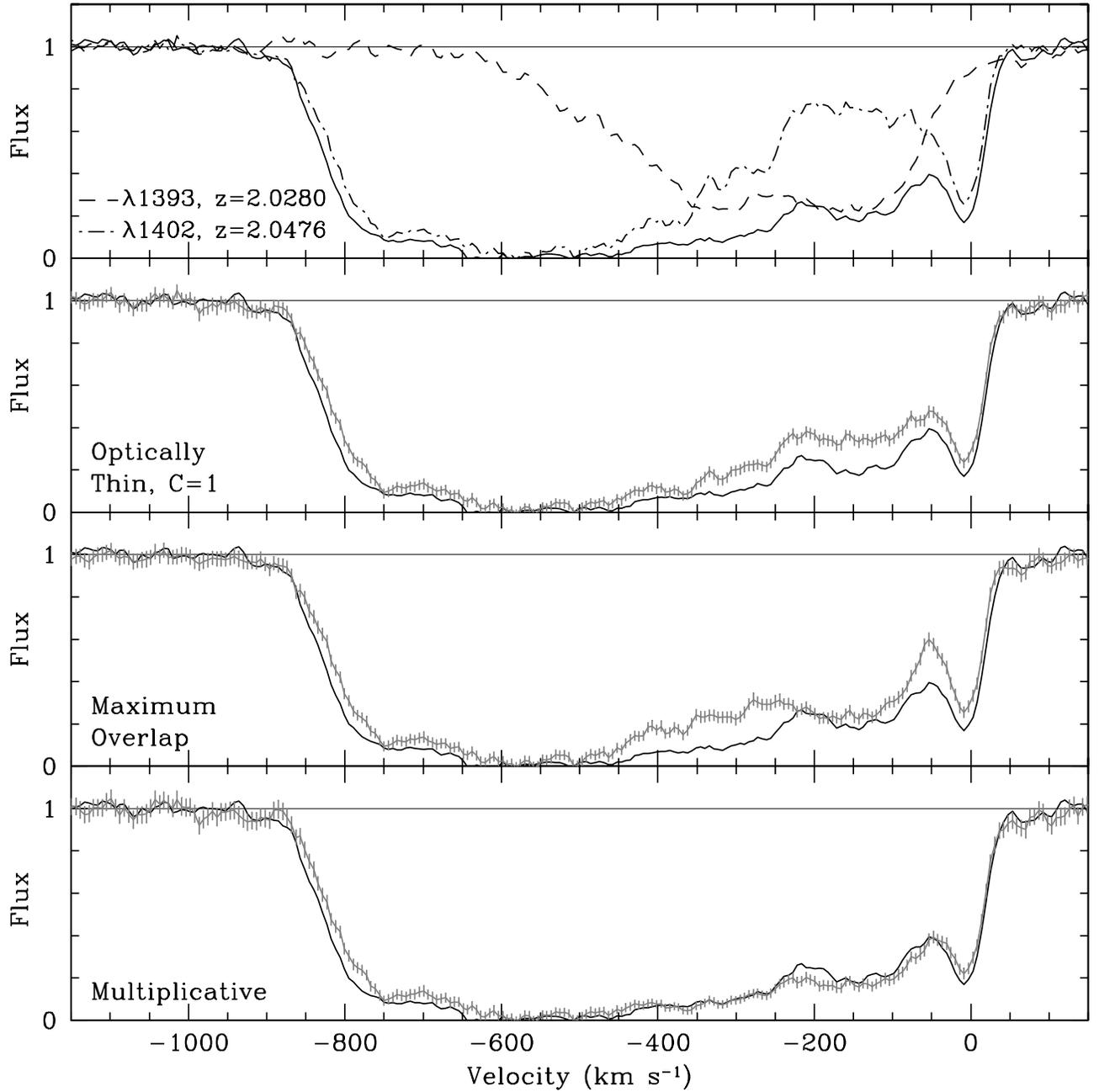}
\caption{Fits to the blended \ion{Si}{4} trough.  The trough containing blended
absorption from both redshift systems is shown as the solid line in all panels.
The fits are shown as lighter lines with {\em total} error bars that include the
observed errors on the flux in the blended trough, so that at each pixel the 
deviation between the actual trough and the fit can be directly compared
to the total accompanying uncertainty.
{\bf Top panel:} all three observed \ion{Si}{4} troughs are overplotted.
The dashed line shows the unblended trough $\lambda$1393 trough, plotted in the $z=2.0280$ frame.
The dot-dashed line shows the unblended trough $\lambda$1402 trough, plotted in the $z=2.0476$ frame.
{\bf Second panel:} the fit and errors shown are for an optically thin
lower-redshift system.
{\bf Third panel:} the fit and errors shown are for an optically thick
lower-redshift system with maximum overlap in covering factor with the
optically thick higher-redshift system.
{\bf Bottom panel:} the fit and errors shown are for the case where each
system's covering fraction describes its fractional coverage of the other
absorption system, so that the residual flux from both optically thick troughs
can be multiplied together to give the residual flux in this blended trough.
}
\label{fig:norm2}
\end{figure}

For an unblended doublet, at each velocity $v$ the normalized residual 
intensities $I_{1}$ and $I_{2}$ (in the stronger and weaker lines, respectively)
can be related to the optical depth in the stronger transition $\tau$ and the
fraction of the emitting source covered by the absorber along our line of sight,
the covering factor $C$ \nocite{sb2}(e.g., {Hall} {et~al.} 2003):
\begin{equation}
 I_{1}(v)=1-C_v(1-e^{-\tau_v})
 \label{eq:I1}
\end{equation}
\begin{equation}
 I_{2}(v)=1-C_v(1- e^{-R\tau_v})
 \label{eq:I2}
\end{equation}
where $R$ measures the relative optical depths of the lines.
For the \ion{Si}{4} $\lambda\lambda 1393,1402$ doublet, $R=0.5$.
In each absorption system we have only one unblended component,
but it can still be used to model the other component.
(For comparison, the two unblended troughs are overplotted 
on the blended trough in the top panel of Figure~\ref{fig:norm2}.)

First we test whether system B can be optically thin, with $C_v=1$.
Using this assumption and equations \ref{eq:I1} and \ref{eq:I2},
the optical depth $\tau_v(\lambda1402,B)$ was
calculated from the observed trough of \ion{Si}{4} $\lambda$1393 in system B.
The blended trough profile in this model should be 
$\exp[-\tau_v(\lambda1402,B)]$
times the profile of \ion{Si}{4} $\lambda$1393 in system A+A$'$.
(The latter profile is taken as identical to the $\lambda$1402 trough profile 
at $z=2.0476$ since system A+A$'$ is optically thick.)
The resulting model blended-trough profile is compared to the 
observed blended-trough profile in the second panel of Figure~\ref{fig:norm2}.
Optically thin absorption from system B
falls short of explaining the depth of the blended trough.

Next we test whether system B can be extremely optically thick, so that
the depth of its absorption is determined only by $C_v$.  In this case,
we have two absorption systems absorbing at each $v$, but with different $C_v$.
The total absorption 
is determined by $C_{v,blended}$, which
depends on what parts of the emitting source(s) are covered
by neither absorption system, by just one, or by both.  
That is, the total absorption depends on the extent to which the two systems
overlap transverse to our line of sight and cover the same parts of the source.
We can rule out the limit of minimum overlap, which yields maximum coverage of
the source: $C_{v,blended}=\min(C_A+C_B,1)$.  In that case
$C_A+C_B>1$ at all $v$, but we do not observed $C_{v,blended}=1$ at all $v$.
Another limiting case is maximum overlap of the absorption systems, which
minimumizes the source coverage: $C_{v,blended}=\max(C_A,C_B)$.
The results of that model are shown in the third panel of
Figure~\ref{fig:norm2}.
It is not an improvement over the optically thin model.
However, at almost all velocities the maximum-overlap model has more residual
flux than seen in the data, while the minimum-overlap model has less.
Thus, overlap in $C_v$ which is less than the maximum possible
by a velocity-dependent amount can explain the data.
Such spatially-distinct, velocity-dependent partial covering
has been seen before in other quasars
(see the Appendix to \nocite{sb2}{Hall} {et~al.} 2003).

The last case we consider is one where each covering fraction describes
the fractional coverage of the other absorption system as well as of the
continuum source, so that $I_{v,blended}=I_AI_B$
and $C_{v,blended}=C_A+C_B-C_AC_B$ (this is case 3 of \nocite{rvs05}{Rupke} {et~al.} 2005).
The results of this model are shown in the bottom panel of
Figure~\ref{fig:norm2}, again assuming A and B are both very optically thick.
The model reproduces the data reasonably well at almost all velocities,
and much more closely overall than the other models considered.

\begin{figure}
\epsscale{1.2}
\plotone{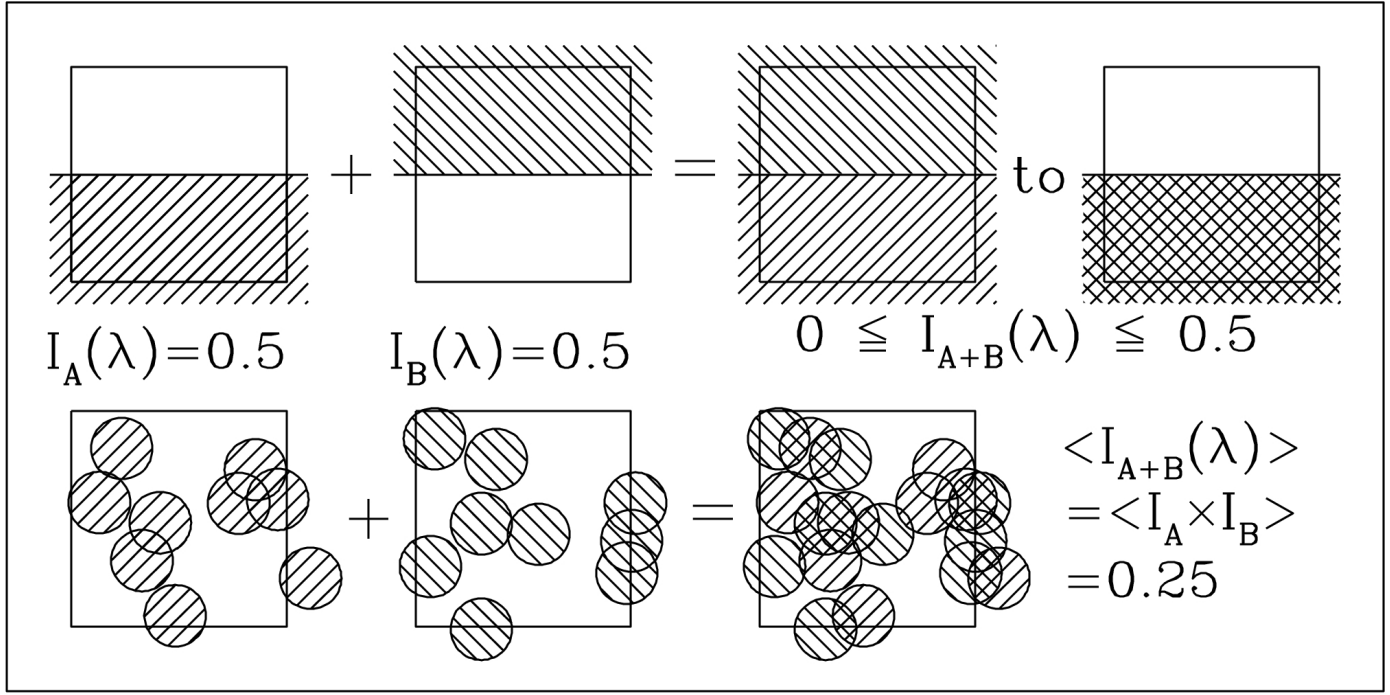}
\caption{An illustration of how observations of optically thick absorption 
systems overlapping in velocity space can constrain absorber substructure.  
Each square is a schematic depiction of a quasar's emission region
at the same wavelength $\lambda$.
The hatched zones are the areas of the emission region covered by an absorber.
The leftmost column depicts absorption by system A only, which is assumed to
produce a normalized residual intensity of $I_A(\lambda)=0.5$ at the wavelength
shown.  The value of $I_A(\lambda)$ is the same regardless of whether the 
absorption is due
to a monolithic flow (top row) or due to randomly placed subunits (bottom row),
here shown as spheres in projection.
Similarly, the second column from the left depicts absorption by system B only,
which is also assumed to produce a normalized residual intensity of 
$I_B(\lambda)=0.5$ at the wavelength shown.
%
When systems A and B overlap, monolithic flows can produce any normalized 
residual intensity in the range $0\leq I_{A+B}(\lambda) \leq 0.5$ (top row, right) 
depending on the specific areas covered by each flow at that wavelength 
$\lambda$.  On the other hand, flows consisting of many randomly placed 
subunits will naturally yield an average value of 
$<$$I_{A+B}(\lambda)$$>$$=$$<$$I_A(\lambda)\times I_B(\lambda)$$>$$=0.25$
{\em at every wavelength $\lambda$ where the systems overlap}
(bottom row, right).
Deviations from this average will occur due to statistical fluctuations,
which will be smaller the more subunits there are.
Measuring the deviations thus enables us to constrain the size and
number of the subunits (see \S 4.1.1).
}
\label{fig:paper}
\end{figure}

The good fit of this model implies that the absorption in one or both of the
systems is produced by many small subunits scattered all over the continuum
source from our point of view.  In that case, the amount of light transmitted
through both systems will naturally be $I_A(v)\times I_B(v)$ 
{\em at every velocity} $v$ (Figure~\ref{fig:paper}).
Deviations will only occur due to statistical 
fluctuations, which will be greater the fewer subunits there are.
It is more difficult, though still possible, to explain the observations using
two `monolithic' systems; that is, systems in which absorption from the ion
in question arises in a single structure along our line of sight spanning the
range of velocities seen in the trough, but with physical coverage of the
source which varies with velocity (e.g., Figure 10 of \nocite{aea99}{Arav} {et~al.} 1999).
Two monolithic flows with unblended residual intensities $I_A(v)$ and $I_B(v)$
can produce any blended residual intensity from 0 to min($I_A(v),I_B(v)$)
essentially independently at each velocity $v$ (Figure~\ref{fig:paper}). 
Thus, two monolithic flows can explain the observations, but only if they just
happen to overlap as a function of velocity in such a way as to mimic the
overlap of two systems of clouds.  
Such an explanation is rather contrived, and we conclude instead that many
small subunits exist in one or both absorption systems.  This conclusion
should of course be tested with observations of additional overlapping
absorption systems in other quasars, to ensure this case is not a fluke.

Note that we have not considered the effects of different covering factors for
the continuum source and broad emission line region.  As seen in 
Figure \ref{fig:lambda}, line emission is a 10\% effect at best, 
and is not a factor at all in the \ion{Si}{4}\,$\lambda$1393 trough of system B.

\subsubsection{Constraints on the Outflow Subunits}

The results above suggest that the absorbers A and B are composed of a number of
optically thick subunits.
We now discuss what we can infer about the parameters of these subunits, in the
limit that each subunit is so optically thick it can be treated as opaque.

Assume that absorber A's residual intensity at some velocity, $I_A(v)$,
is created by $N_A$ subunits intercepting our line of sight, and similarly for 
absorber B.  When the two absorbers overlap along the line of sight, there will
be $N=N_A+N_B$ subunits along the line of sight.  The average transmitted flux
$i$ in this case will be $<i>=(1-p)^N$, where $p$ is the average fraction 
of the quasar's emission covered by an individual subunit.


If an average $N$ over all velocities is well defined,
the pixel-to-pixel
variations around the average value $<i>$ will be distributed with variance
$\sigma^2=\sigma_I^2+\sigma_i^2$, where $\sigma_I$ is the instrumental error
and $\sigma_i$ is given by
\begin{equation}
\sigma_i^2 = \sigma_{intrinsic}^2
+(1-p)^{2N}\left(\frac{N^2\sigma_p^2}{(1-p)^2} + [\ln(1-p)]^2\sigma_N^2\right).
\end{equation}
For example, fixed $N$ at all velocities would have $\sigma_N^2=0$, while
a Poisson distribution with an average of $N$ would have $\sigma_N^2=N$.
The intrinsic variance at fixed $N$ and $p$, $\sigma_{intrinsic}^2$, 
is caused by the random overlap (or lack thereof) of $N$ subunits of
uniform projected fractional area $a$.
The relation between $p$ and $a$, and the form of $\sigma_{intrinsic}^2$,
depends on the shape of the subunits and of the quasar's emitting region.  
In the Appendix we give formulae for the
cases of rectangular subunits of width $a$ and unit length and of circular
subunits of area $a$, under the approximation that the emitting region of the
quasar is projected on the sky as a square of unit area and uniform surface 
brightness (see the discussion in the Appendix).  
In both cases, $\sigma_p^2 \propto \sigma_a^2$.  If $\sigma_a$ is negligible,
there are two unknowns ($a$ and $N$) and two observables ($<i>$ and $\sigma$)
which can be used to solve for them.  

More generally, we can constrain the subunit number and size as follows.
We have a predicted profile $i(v)=I_AI_B$ and an observed profile $I(v)$,
both of which depend on velocity.
In our case, the wide range of $i$ over the full trough 
and the smooth pixel-to-pixel distribution of $i$
cannot simultaneously be reproduced at fixed $N$.
Reproducing the wide range of $i$ would require a small $N$,
which would not generate as smooth a velocity profile as observed.  
Each subunit will probably have a velocity dispersion of 
only $\sim$10\ \kms\ \nocite{pet97}({Peterson} 1997),
so for small $N$ strong variations in $i$ would be seen on that velocity scale.
Thus, the range in $i$ means either $N$ or $a$ varies with velocity, or both 
do.  To simplify the problem, we assume the subunits have a uniform size so that
$a$ is constant and $\sigma_a = 0$.  (This should be an adequate approximation
if the subunits have a characteristic size scale.)
If we then assume a value for $a$, we can calculate 
a predicted $N$ for each pixel as $N=\log i / \log (1-p)$,
using the expression for $p(a)$ appropriate to the chosen geometry.
The observed profile $I$ differs slightly from the predicted profile $i=I_AI_B$,
due to the intrinsic variance on the total covering factor of $N$ clouds
($\sigma_{intrinsic}^2$)
and to the errors on $I_A$ and $I_B$ ($\sigma_A$ and $\sigma_B$, respectively).
Setting $\sigma_p\propto \sigma_a=0$ as discussed above and approximating 
the variance on $N$ as $\sigma_N^2=N$, we have
\begin{equation}
\sigma_i^2 \simeq \sigma_{intrinsic}^2
+(1-p)^{2N}N[\ln(1-p)]^2 + I_B^2\sigma_A^2+I_A^2\sigma_B^2.
\end{equation}
The probability of observing a residual intensity $I\pm\sigma_I$ 
in a pixel, given a predicted value $i$ and associated $\sigma_i$, is
\begin{equation}
P(I\pm\sigma_I|i\pm\sigma_i) = \frac{1}{\sqrt{2\pi(\sigma_I^2+\sigma_i^2)}}\exp\left[-\frac{(I-i)^2}{2(\sigma_I^2+\sigma_i^2)}\right].
\end{equation}
Each pixel has a different $\sigma_i$ which depends on the adopted $a$.
To choose the best model, we find the value of $a$ that maximizes the likelihood
of the observations: $L=\prod_k P(I_k\pm\sigma_{I_k}|i_k\pm\sigma_{i_k})$.
Note that a systematic error in $I$ (e.g., due to a continuum estimate which is
too high or too low) will yield a systematic error in $a$.

We use the velocity range $-700<v<-75$~\kms\ to calculate $L$, avoiding both the
narrow system A$'$ and the high-velocity edge of the trough from system A where
convolution with the instrumental line spread function may alter the true
relative absorption depths in the two lines of a doublet \nocite{gecc99}({Ganguly} {et~al.} 1999).
We find a best-fit relative filament width $w=0.0135$,
with a 99.994\% (4$\sigma$) probability range of $0.0014<w<0.0430$.
We find a best-fit relative cloud radius $r=0.081$,
with a 99.994\% (4$\sigma$) probability range of $0.029<r<0.143$.
There is no statistically significant difference between the likelihood 
of the two fits.

To convert these to physical sizes, we model the quasar's emission as being 
from a \nocite{ss73}{Shakura} \& {Sunyaev} (1973) accretion disk with viscosity
parameter $\alpha=0.1$ radiating at the Eddington limit.  (We discuss the issue
of coverage of the quasar's broad emission line region at the end of the 
section.) For this quasar we estimate $M_{BH}=6.2 \times 10^8$ M$_\odot$ from
the second moment of its \ion{Mg}{2} emission line and its 3000~\AA\ continuum
luminosity, using the methods of Rafiee et al. (2007, in preparation).
For those parameters, 99\% of the continuum emission at rest-frame 1400\,\AA\ 
comes from $r<150R_{Sch}$, where $R_{Sch}=2GM_{BH}/c^2=1.8\times10^{14}$~cm
is the Schwarzschild radius of the black hole.
Since the relative sizes derived above were referenced to a square, not a 
circle, we adopt the square that has the same area as a circle with radius
$150R_{Sch}$, which has sides of length $l=4.8\times10^{16}$~cm.
Thus, we find a best-fit filament width of $w=6.5\times10^{14}$~cm,
with a 4$\sigma$ range of $6.7\times10^{13}<w<2.1\times10^{15}$~cm,
and a best-fit cloud radius $r=3.9\times10^{15}$~cm, with a 4$\sigma$ range
of $1.4\times10^{15}<r<6.9\times10^{15}$~cm.

These sizes, small on astronomical scales, suggest an origin for the subunits
in the accretion disk for either geometry.  A plausible length scale for 
structures originating in an accretion disk is the scale height $h$ of its 
atmosphere (Equation 2.28 of \nocite{ss73}{Shakura} \& {Sunyaev} 1973).\footnote{If the accretion disk
has a strong magnetic field, the pressure scale height may be a less plausible 
characteristic length.  Numerical simulations of accretion
disks do not yet conclusively show if another characteristic scale is produced
by magnetohydrodynamic turbulence \nocite{arm04}({Armitage} 2004).}
At large radii, 
$h\simeq 3R^3kT_s/4GM_{BH}m_pz_0$ where $R$ is the distance from the black hole,
$T_s$ is the disk surface temperature
and $z_0$ is the disk half-thickness.
(Though not obvious from the above, $h<z_0$ because the
disk surface temperature is lower than its midplane temperature.)
In this object, the best-fit filament width equals the scale height $h$ at
$r=5500R_{Sch}=9.9\times 10^{17}$~cm
and the best-fit cloud radius equals the scale height $h$ at
$r=25000R_{Sch}=4.5\times 10^{18}$~cm. 
The various parameters for our two geometries are summarized in Table 2.

Strikingly, the first of those distances from the central source is
equal to the distance the absorber must have to cover the
emission from the quasar's broad emission line region (BELR).
As seen in Figure 4, the line emission in the region of the absorption troughs
reaches at most 10\% of the continuum level, and at least system A covers both
the continuum emission region and the \ion{Si}{4}/\ion{O}{4}] BELR.  In other
transitions, both systems at least partially cover the \ion{N}{5} and \ion{C}{4}
BELRs, and at least system A covers the \ion{Mg}{2} BELR.  Since AGN BELRs are
stratified, with lower-ionization gas located farther from the quasar, to be
conservative we assume both systems lie exterior to the \ion{Mg}{2} BELR in \jo.
We use a relationship between $L_\lambda$(3000\AA) and $R_{\rm BELR,MgII}$ 
derived from reverberation-mapping data (\nocite{pea04}{Peterson} {et~al.} 2004; Rafiee et al. 2007,
in preparation) to obtain $R_{\rm BELR,MgII}=9.1 \times 10^{17}\ {\rm cm} =5000
R_{Sch}$ for \jo.  Given the $\pm$25\% 1$\sigma$ scatter in this relationship, 
this distance is in excellent agreement with the distance required for 
filamentary absorber subunits to have widths matching the disk scale height.
Of course, the absorber could be located at any $R>R_{\rm BELR,MgII}$,
so spherical clouds of size equal to the disk scale height could still
match the data if the outflow arises at sufficiently large radii.

We have outlined a consistent picture wherein systems A and B, 
whether they consist of opaque filaments or clouds, are launched from 
the accretion disk exterior to the \ion{Mg}{2} BELR with a subunit size
comparable to the scale height of the accretion disk atmosphere at that radius.
As a system accelerates, its typical density will decrease and its typical
ionization will increase, explaining the presence of high ionization
species in flows arising from a low-ionization emission-line region.
When the systems cross our line of sight, they have line-of-sight velocities of
$v_{los}=-2000$\,\kms\ for system A and $v_{los}=-3600$\,\kms\ for system B.
For System A, $|v_{los}|$ is comparable to the $v_{orbital}=2900$\,\kms\ 
expected at its inferred launch radius of 5500$R_{Sch}$.
For System B, $|v_{los}|$ is larger than the $v_{orbital}=1400$\,\kms\ 
expected at its inferred launch radius of 25000$R_{Sch}$.
The spherical cloud dispersal time would be of order $\sim$110 years for
$T\sim 10^4$\,K, so the subunits will not disperse on their own between launch
and crossing our line of sight.  However, partial shadowing of a subunit will
produce differential radiative acceleration of the subunit.  Substantial 
radiative acceleration could thus shorten the subunit lifetimes considerably.

One potential complication is that the observed profile of the overlapping
trough deviates from the multiplicative prediction (Figure 6, bottom panel)
in a manner that is not random on velocity scales larger than $\sim$10\ \kms. 
However, deviations on such scales should be random if, as expected, 
the individual subunits have velocity dispersions of that order.
Instead, the deviations seem to be coherent on $\sim$100\ \kms\ scales.
It may be that the subunits do have velocity widths of that order
due to microturbulence (\nocite{bfbk00}{Bottorff} {et~al.} 2000).
Another possible explanation is that the outflow consists of filaments
wherein the material is accelerated so that its line-of-sight
velocity increases by $\sim$100\ \kms\ as it crosses the line of sight
(e.g., \nocite{aea99}{Arav} {et~al.} 1999).
Deviations from the expected profile should then persist for $\sim$100\ \kms\ 
instead of $\sim$10\ \kms.  As compared to a model without line-of-sight 
acceleration, there could be the same average number of filaments, but the
number would change more slowly with velocity (although other effects, 
such as filaments not being exactly parallel, can affect that as well).
Observations of additional overlapping systems would be useful 
for investigating this issue.

We note that \nocite{goo03}{Goodman} (2003) have shown that thin accretion 
disks without winds will be unstable to self-gravity beyond 
$r_{Q=1} \simeq 2740 (10^8\alpha l_E^2/M_{BH})^{2/9} R_{Sch}$
where $l_E$ is the Eddington ratio;
using the parameters adopted herein, \jo\ has $r_{Q=1} \simeq 1100R_{Sch}$.
However, removal of angular momentum by a disk wind might help stabilize a thin
disk (\S 4.3 of \nocite{goo03}{Goodman} 2003), and there is reason to believe
such a process operates in AGN.  Reverberation mapping places the BELRs of many
AGN at $r>r_{Q=1}$, and there is evidence that BELRs are flattened 
\nocite{vwb00,sea05,aa05}({Vestergaard}, {Wilkes}, \&  {Barthel} 2000; {Smith} {et~al.} 2005; {Aars} {et~al.} 2005) 
as expected if they are located at the bases of accretion disk winds 
\nocite{mur95}({Murray} {et~al.} 1995).  Furthermore, quasar spectral energy
distributions are consistent with marginally gravitationally stable disks
extending out to $\sim 10^5R_{Sch}$ \nocite{sg03}({Sirko} \& {Goodman} 2003). 

Lastly, we note that
there is no contradiction in using the continuum source size to derive the scale
size of the subunits for an outflow the size of the BELR.  This is because the
continuum source has a surface brightness $\simeq 2100$ times that of the BELR.
That number is the ratio of the continuum flux near 1400\,\AA\ in \jo\ to the
\ion{Si}{4}/\ion{O}{4}] flux, which we take to be $\simeq 9$,
times the ratio of the areas of the \ion{Si}{4}/\ion{O}{4}] 
BELR and the 1400\,\AA\ continuum source.\footnote{The size of the 
\ion{Si}{4}/\ion{O}{4}] BELR has been measured in only three AGN \nocite{pea04}({Peterson} {et~al.} 2004).
On average, it is comparable in size to the \ion{C}{4} BELR.  We therefore use
the relationship between $L_\lambda$(1350\AA) and $R_{\rm BELR,CIV}$ given by
\nocite{pee06}{Peterson} {et~al.} (2006) to derive $R_{\rm BELR,SiIV}=4.1 \times 10^{17}$~cm for \jo.}
If $N$ subunits of the absorber each cover
a fractional area $a$ of the continuum source, $Nx$ subunits of the absorber 
will each cover a fractional area $a/x$ of the BELR. For large $N$ and small $a$
the residual intensity of each region is equal, $i=(1-a)^N\simeq(1-a/x)^{Nx}$,
but the variance on $i$ from the BELR will be a factor $\simeq0.1/x$ smaller
than the variance on $i$ from the continuum source.
Thus, an absorber covering both the continuum source and BELR will have 
essentially the same residual intensity $i$ and variance $\sigma_i^2$ 
(used to derive the absorber size constraints via Equation 6)
as an absorber covering only the continuum source.

\subsection{Possible \ion{Si}{4} Line-Locking}

We now return to the issue of whether systems A+A$'$ and B can be line-locked.  
Line-locking occurs when the reduction in line-driving flux caused by the
shadow of one system decelerates the other, shadowed system 
so that two systems end up with the same acceleration (which may be nonzero).
The two systems thereafter maintain
a constant velocity separation that keeps one system shadowed
\nocite{bm89}({Braun} \& {Milgrom} 1989).  (However, there is some debate in the literature as to whether
line-driven winds are unstable to the growth of shocks \nocite{ocr88,poht04}({Owocki}, {Castor}, \& {Rybicki} 1988; {Pereyra} {et~al.} 2004).
If shocks can develop, they could
accelerate the wind
out of an otherwise stable line-locking configuration.)
For line-locking to occur in an accelerating flow, 
there are two possibilities.
System B could have appeared 
along a sightline linking the continuum source and system A+A$'$
at $2.0280<z<2.0476$ and accelerated
until it reached $z=2.0280$ and overlapped system A+A$'$ at $z=2.0476$.
Alternatively, system A+A$'$ could have appeared at $z>2.0476$ and
accelerated until it reached $z=2.02476$ and overlapped system B at $z=2.0280$.

The latter scenario can be ruled out because the greatest deceleration of
system A+A$'$ would have occurred before it reached $z=2.0476$,
when it was shadowed by the deepest part of system B.
Instead, the deepest part of system B is observed to be shadowed
by the shallowest part of system A.
If line-locking was going to occur in this scenario it would have had to set in 
when the shadowing was greatest (or earlier than that, if less than full 
shadowing produced sufficient deceleration).  If it did not happen then,
it will not happen with the observed, lesser amount of shadowing.

The former scenario of an accelerating system B which has ended up
line-locked is plausible.  The observed shadowing as a function of velocity 
could in principle
have halted system B.

One requirement of this former scenario, however, is that the narrow absorption
at $z=2.0476$ (system A$'$) should not be associated with system A, 
the broad absorption immediately shortwards of it.
If they were associated, then some of the gas in system B at 
$-350<z<-50$ \kms\ should have come to a halt at 0 \kms, where the shadowing
by system A$'$ would have been greater than the current shadowing by system A.
System A$'$ must be located farther from the quasar than either system A or B,
in this scenario.

The optically thickest part of system A
is likely at $-650<v<-450$ \kms, where numerous low-ionization species are seen.
If any gas in system B was observed at $v<-650$ \kms, that gas
would have passed the point of maximum shadowing without becoming line-locked.
In fact, no gas in system B is seen at $v<-650$ \kms,
consistent with system B being line-locked.
One argument against this scenario is that if system B has been halted by the
observed shadowing, gas at different velocities in that system has been halted
by different amounts of shadowing.  For example, gas at $-200$ \kms\ 
has been halted by shadowing of only $\sim$30\% of the continuum, while 
gas at $-450$ \kms\ has been halted by shadowing of $\sim$95\% of the continuum.
It may be more physically plausible to suggest that gas at $-450$ \kms\ has
been halted, but that gas at $-200$ \kms\ has not yet been sufficiently shadowed
to become line-locked.  In other words,
in this model system B is in the process of becoming line-locked.  
However, comparison of the SDSS and UVES spectra shows no evidence for 
variability in these \ion{Si}{4} troughs.  The timescale for velocity changes 
in this scenario could be longer than 1.4 years (rest-frame),
which would rule out line locking in a \nocite{mur95}{Murray} {et~al.} (1995) disk wind 
in which the entire acceleration phase lasts $\sim1.6$~years,
or the line-locking could be occuring in a helical flow, stable on timescales
of years, in which
our sightline intercepts the flow before the gas becomes line-locked.

Finally, note that the \ion{Si}{4} profiles in \jo\ are intriguingly similar to
some of the potentially line-locked \ion{N}{5} profiles seen in RX~J1230.8+0115 
\nocite{gmcs03}({Ganguly} {et~al.} 2003).  The $z=0.1058$ system in that object has a profile similar to 
that of system A+A$'$ (strongest absorption at both ends of the profile),
and its $z=0.1093$ system is similar to that of system B (optically thick,
with the strongest absorption in the middle of the profile, at a velocity
corresponding to the weakest absorption in the other system).
Both systems have only about half the velocity widths of those in \jo,
however, and the relative velocities of the two systems are reversed ---
the weaker, single-peaked absorption profile has the lower outflow velocity.
It is also worth noting that the Ly$\alpha$ absorption profile in each object
appears to share the same covering factor as the species discussed above,
while at least one moderately higher-ionization species in each object
(\ion{N}{5} here, and \ion{O}{6} in RX~J1230.8+0115)
has a larger covering factor which yields nearly black absorption troughs.
Whether these similarities are just coincidences 
will require data on more candidate line-locking systems.
(The line-locked systems in Q~1511+091 studied by \nocite{splh02}{Srianand} {et~al.} (2002) are much more
complex, but do not seem to include any profiles similar to those in \jo.)

\section{Conclusions}\label{end}

We find that the \ion{C}{4} BAL trough at $z=1.87988$ in the spectrum of \jo\ 
($v=-18400$\ \kms\ relative to the quasar's rest frame) has likely undergone
an acceleration of $a = 0.154 \pm 0.025 \mbox{\ cm\ s}^{-2}$ over a period of 
1.39 rest-frame years.  This is the largest acceleration yet reported 
in a BAL trough $\geq$1000\ \kms\ wide.

We also derive constraints on the outflow properties of two absorption systems, 
overlapping and possibly line-locked in \ion{Si}{4}, 
at $z=2.0420$ and $z=2.0254$
($v=-2000$\ \kms\ and $v=-3600$\ \kms\ relative to the quasar, respectively).
The overlapping trough in common to both systems indicates that
at least one of the systems must consist of individual subunits.
This contrasts with results strongly suggesting that the BELR itself consists of
a smooth flow, rather than a clumped one (\nocite{lbhf06}{Laor} {et~al.} 2006),
but agrees with results for a narrow intrinsic absorber in the gravitational
lens RXS J1131$-$1231 (\nocite{schs07}{Sluse} {et~al.} 2007).

Assuming identical, opaque subunits, our data are consistent with
spherical clouds of radius $r\simeq 3.9\times 10^{15}$~cm
or linear filaments of width $w\simeq 6.5\times 10^{14}$~cm.
These subunits must be located at or beyond the \ion{Mg}{2} broad emission line
region.
At that distance, the above filament width is equal to the
predicted scale height of the outer atmosphere of a thin accretion disk.
Insofar as that is
a natural length scale for structures originating in an accretion disk,
these observations are evidence that the accretion disk is the source of 
the absorption systems.
%
It would be useful to obtain high-resolution spectra of additional cases 
of distinct but overlapping intrinsic absorption troughs in quasar spectra
to determine if this case is representative.
If so, it would also be worth extending this work's analytic study of 
the implications of the residual intensity variance to numerical studies 
including a realistic quasar geometry, a range in absorber sizes
and optical depths, etc.

\acknowledgments
We thank N. Murray for discussions, and the referee for helpful comments.
P. B. H. is supported by NSERC, and S. I. S. was supported by an NSERC 
Undergraduate Summer Research Assistantship.
The SDSS and SDSS-II (http://www.sdss.org/) are funded by the Alfred P. Sloan
Foundation, the Participating Institutions, the National Science Foundation,
the U.S. Department of Energy, NASA, the Japanese Monbukagakusho, 
the Max Planck Society, and the Higher Education Funding Council for England,
and managed by the Astrophysical Research Consortium for the Participating Institutions: American Museum of Natural History, Astrophysical Institute Potsdam, University of Basel, Cambridge University, Case Western Reserve University, University of Chicago, Drexel University, Fermilab, the Institute for Advanced Study, the Japan Participation Group, Johns Hopkins University, the Joint Institute for Nuclear Astrophysics, the Kavli Institute for Particle Astrophysics and Cosmology, the Korean Scientist Group, the Chinese Academy of Sciences,
Los Alamos National Laboratory, the Max-Planck-Institute for Astronomy,
the Max-Planck-Institute for Astrophysics,
New Mexico State University, Ohio State University, University of Pittsburgh, University of Portsmouth, Princeton University, the United States Naval Observatory, and the University of Washington.

\appendix

Consider the case of an absorber consisting of opaque subunits
of a uniform shape.
Suppose our line of sight to a quasar's emitting regions is intercepted by $N$
of these subunits, randomly distributed transverse to the line of sight.
Then the scatter possible in the covering fraction at fixed $N$ due to the
random overlap (or lack thereof) of the subunits with each other will depend on
the shape of the subunits.  To obtain expressions for this variance, we 
approximate the quasar's emitting regions as a square of uniform surface 
brightness on the plane of the sky.  
We do this solely because expressions for the variance
have been derived for the case of the unit square covered by two relevant
subunit geometries: circles of area $a$ and filaments of unit length and 
width $a$.
We take the first case to represent a true cloud model, and the second to
represent a magnetically confined `filament' model.

The case of the unit square randomly overlapped by filaments parallel to each
other and to two sides of the square, and of unit length and width $a$, is 
treated by \nocite{gplin}Robbins (1944).  The unit square is defined as the set of points
\{$0\leq x\leq 1;0 \leq y \leq 1$\}.  The filaments that overlap the square
are centered at $y=0.5$ and distributed randomly in $x$
over $-\frac{a}{2}\leq x\leq 1+\frac{a}{2}$.  Because of edge effects, the 
average area covered by a filament is $p=\frac{a}{1+a}$, and the average
area uncovered by $N$ filaments is $i=(1-p)^N$.
The variance in the fractional area covered is
\begin{eqnarray}
\sigma_{\rm filaments}^2=(1-a)^2(1-2p)^N -(1-p)^{2N} 
+\frac{2a[(1-p)^{N+1}-(1-a)(1-2p)^{N+1}]}{(N+1)p}\nonumber\\
-\frac{2a^2[(1-p)^{N+2}-(1-2p)^{N+2}]}{(N+1)(N+2)p^2}
\end{eqnarray}
for $a<0.5$.

In the case of the unit square randomly overlapped by circles of area $a$,
circles that overlap the square are distributed such that their centers 
are within a distance $r=\sqrt{a/\pi}$ of the unit square.
Again the average area uncovered by $N$ circles is given by $i=(1-p)^N$,
but in this case $p=\pi r^2/(1+4r+\pi r^2)$. 
The variance in the fractional area covered can be derived from expressions
given by \nocite{galp}{Kendall} \& {Moran} (1963), yielding
\begin{eqnarray}
\sigma_{\rm circles}^2 =
\left[ \frac{1+4r-\pi r^2}{1+4r+\pi r^2} \right]^N
\left(1 -4\pi r^2 +\frac{64}{3}r^3 -8r^4 \right)
-\left(\frac{1+4r}{1+4r+\pi r^2}\right)^{2N}\nonumber\\
+2 \int_0^{2r}\left[1-\frac{2r^2 (\pi -\cos^{-1}\frac{q}{2r} +\frac{q}{2r}\sin\left(\cos^{-1}\frac{q}{2r}\right)}{1+4r+\pi r^2}\right]^N
(\pi q -4q^2 + q^3) ~dq
\end{eqnarray}
for $a<0.5$.  The integral must be evaluated numerically for most $N$.

For the same $a$ and $N$, $\sigma_{\rm circles}^2 > \sigma_{\rm filaments}^2$.
This can be understood by placing a subunit of either type in the center of the
square
and considering the probability that a second subunit of the same type 
will overlap the first.
There is an area $2a$ in which a second filament can be placed to have 
some overlap with the first (filament centers at $0.5-a<x<0.5+a$). 
There is an area $4a$ in which a second circle can be placed to have 
some overlap with the first (circles centered within $2\sqrt{a\over\pi}$
of $\{0.5,0.5\}$, for an area of $\pi (2\sqrt{a\over\pi})^2 = 4a$).
If $a$ is small,
the most likely value of $i$ 
is $i=1-2a$ for both geometries, but with circles there is a higher probability
of $i>1-2a$ and thus a larger variance.

\begin{deluxetable}{ccccrcc}
\tablecaption{SDSS J0242+0049 Spectroscopic Observations and Inferences\label{spec}}
\tablewidth{480pt}
\tablehead{
\colhead{}& \colhead{SDSS}& \colhead{SDSS}& \colhead{Epoch}& \colhead{$\Delta t_{rest}$}& \colhead{\ion{Si}{4}, \ion{C}{4} Shift}& \colhead{\ion{Si}{4}, \ion{C}{4} Shift}\\[0.2ex]
\colhead{Source}& \colhead{Plate}& \colhead{Fiber}& \colhead{in MJD}& \colhead{(days)}& \colhead{vs. MJD 52188}& \colhead{vs. MJD 53619}
}
\startdata 
         SDSS~(1)& 408& 576&   51821& $-$80& 0, 0& 1, 4\\
         SDSS~(2)& 707& 332&   52177&    36& --- & --- \\
         SDSS~(3)& 706& 617&   52199&    43& --- & --- \\
  SDSS~Avg.~(2+3)& ---& ---& (52188)&    40& --- & 1, 3\\
SDSS~Avg.~(1+2+3)& ---& ---& (52066)&     0& --- & 1, 3\\
             UVES& ---& ---&   53619&   507& 1, 3& --- \\
\enddata
\tablecomments{Epochs are given on the Modified Julian Day (MJD) system.
The {\em rest-frame} time interval $\Delta t_{rest}$
is given relative to MJD 52066.
Velocity shifts of absorption lines are given in SDSS pixels (69 \kms); 
the \ion{C}{4} shift is the first number and
the \ion{Si}{4} shift is the second number.
}
\end{deluxetable}

\begin{deluxetable}{ccccccc}
\tabletypesize{\scriptsize}
\tablecaption{SDSS J0242+0049 Subunit Parameters\label{model}}
\tablewidth{525pt}
\tablehead{
\colhead{Subunit}& \colhead{Avg. Number}&      \colhead{Best-fit Relative}& \colhead{Relative 99.994\%}& \colhead{Best-fit Physical}&    \colhead{Physical 99.994\%}&     \colhead{Atmospheric}\\[0.2ex]
\colhead{Geometry}& \colhead{of Subunits $\bar{N}$}& \colhead{Width or Radius}&   \colhead{Confidence range}&  \colhead{Width or Radius}& \colhead{Confidence range (cm)}& \colhead{Scale Height Distance} 
}
\startdata 
Filaments& $203\pm81$& 0.0135& $0.0014-0.0430$& $6.5\times10^{14}$~cm& $6.7\times10^{13}-2.1\times10^{15}$& $9.9\times10^{17}$~cm = 5500~$R_{Sch}$\\
Spheres&   $177\pm71$&  0.081& $0.029-0.143$&   $3.9\times10^{15}$~cm& $1.4\times10^{15}-6.9\times10^{15}$& $4.5\times10^{18}$~cm = 25000~$R_{Sch}$\\
\enddata
\tablecomments{The average number of subunits $\bar{N}$ is the number 
of subunits responsible for absorption at each pixel, averaged over all pixels.
The total number of 
subunits present depends on the unknown velocity width of each subunit.
The atmospheric scale height distance is the distance from the black hole at
which the accretion disk atmospheric scale height equals the best-fit width or
radius of the subunit in question; see \S 4.1.  $R_{Sch}$ refers to the 
Schwarzschild radius of a black hole with mass $6.2 \times 10^8$ M$_\odot$.
}
\end{deluxetable}

\clearpage
\begin{thebibliography}{}

\bibitem[{Aars}, {Hough}, {Yu}, {Linick}, {Beyer},  {Vermeulen}, \& {Readhead} 2005]{aa05}
{Aars}, C.~E., {Hough}, D.~H., {Yu}, L.~H., {Linick}, J.~P., {Beyer}, P.~J.,  {Vermeulen}, R.~C., \& {Readhead}, A.~C.~S. 2005, \aj, 130, 23

\bibitem[{Arav}, {Korista}, {de Kool}, {Junkkarinen}, \&  {Begelman} 1999]{aea99}
{Arav}, N., {Korista}, K.~T., {de Kool}, M., {Junkkarinen}, V.~T., \&  {Begelman}, M.~C. 1999, \apj, 516, 27

\bibitem[{Armitage} 2004]{arm04}
{Armitage}, P.~J. 2004, {Theory of Disk Accretion onto Supermassive Black  Holes} (ASSL Vol.~308: Supermassive Black Holes in the Distant Universe), 89

\bibitem[{Barlow}, {Hamann}, \& {Sargent} 1997]{bhs97}
{Barlow}, T., {Hamann}, F., \& {Sargent}, W. 1997, in ASP Conf. Ser. 128: Mass  Ejection from Active Galactic Nuclei, 13

\bibitem[{Barlow}, {Junkkarinen}, \&  {Burbidge} 1989]{bjb89}
{Barlow}, T., {Junkkarinen}, V., \& {Burbidge}, E. 1989, \apj, 347, 674

\bibitem[{Barlow}, {Junkkarinen}, \&  {Burbidge} 1992a]{bjb92}
{Barlow}, T., {Junkkarinen}, V., \& {Burbidge}, E. 1992a, in  American Astronomical Society Meeting, Vol. 181, 1106

\bibitem[{Barlow}, {Junkkarinen},  {Burbidge}, {Weymann}, {Morris}, \& {Korista} 1992b]{bea92}
{Barlow}, T., {Junkkarinen}, V., {Burbidge}, E., {Weymann}, R., {Morris}, S.,  \& {Korista}, K. 1992b, \apj, 397, 81

\bibitem[{Barlow} 1994]{bar94}
{Barlow}, T.~A. 1994, \pasp, 106, 548

\bibitem[{Bottorff}, {Ferland}, {Baldwin}, \& {Korista} 2000]{bfbk00}
{Bottorff}, M.~C., {Ferland}, G.~J., {Baldwin}, J., {Korista}, K. 2000,
\apj, 542, 644

\bibitem[{Braun} \& {Milgrom} 1989]{bm89}
{Braun}, E. \& {Milgrom}, M. 1989, \apj, 342, 100

\bibitem[{Bromage}, {Boksenberg}, {Clavel}, {Elvius},  {Penston}, {Perola}, {Pettini}, {Snijders}, {Tanzi}, \& {Ulrich} 1985]{bea85}
{Bromage}, G., {Boksenberg}, A., {Clavel}, J., {Elvius}, A., {Penston}, M.,  {Perola}, G., {Pettini}, M., {Snijders}, M., {et al.} 1985, \mnras, 215, 1

\bibitem[{Dekker}, {D'Odorico}, {Kaufer}, {Delabre}, \&  {Kotzlowski} 2000]{uves}
{Dekker}, H., {D'Odorico}, S., {Kaufer}, A., {Delabre}, B., \& {Kotzlowski}, H.  2000, in Proc. SPIE Vol. 4008, Optical and IR Telescope Instrumentation and  Detectors, ed. M.~{Iye} \& A.~{Moorwood}, 534

\bibitem[{Foltz}, {Weymann}, {Morris}, \&  {Turnshek} 1987]{fol87}
{Foltz}, C.~B., {Weymann}, R.~J., {Morris}, S.~L., \& {Turnshek}, D.~A. 1987,  \apj, 317, 450

\bibitem[{Fukugita}, {Ichikawa}, {Gunn}, {Doi},  {Shimasaku}, \& {Schneider} 1996]{fuk96}
{Fukugita}, M., {Ichikawa}, T., {Gunn}, J.~E., {Doi}, M., {Shimasaku}, K., \&  {Schneider}, D.~P. 1996, \aj, 111, 1748

\bibitem[{Gabel}, {Crenshaw}, {Kraemer}, {Brandt},  {George}, {Hamann}, {Kaiser}, {Kaspi}, {Kriss}, {Mathur}, {Mushotzky},  {Nandra}, {Netzer}, {Peterson}, {Shields}, {Turner}, \& {Zheng} 2003]{gea03}
{Gabel}, J.~R., {Crenshaw}, D.~M., {Kraemer}, S.~B., {Brandt}, W.~N., {George},  I.~M., {Hamann}, F.~W., {Kaiser}, M.~E., {Kaspi}, S., {et al.} 2003, \apj, 595, 120

\bibitem[{Gallagher}, {Brandt}, {Wills}, {Charlton},  {Chartas}, \& {Laor} 2004]{gea04}
{Gallagher}, S.~C., {Brandt}, W.~N., {Wills}, B.~J., {Charlton}, J.~C.,  {Chartas}, G., \& {Laor}, A. 2004, \apj, 603, 425

\bibitem[{Gallagher}, {Schmidt}, {Smith}, {Brandt},  {Chartas}, {Hylton}, {Hines}, \& {Brotherton} 2005]{gea05}
{Gallagher}, S.~C., {Schmidt}, G.~D., {Smith}, P.~S., {Brandt}, W.~N.,  {Chartas}, G., {Hylton}, S., {Hines}, D.~C., \& {Brotherton}, M.~S. 2005,  \apj, 633, 71

\bibitem[{Ganguly}, {Charlton}, \&  {Eracleous} 2001]{gce01}
{Ganguly}, R., {Charlton}, J.~C., \& {Eracleous}, M. 2001, \apjl, 556, L7

\bibitem[{Ganguly}, {Eracleous}, {Charlton}, \&  {Churchill} 1999]{gecc99}
{Ganguly}, R., {Eracleous}, M., {Charlton}, J.~C., \& {Churchill}, C.~W. 1999,  \aj, 117, 2594

\bibitem[{Ganguly}, {Masiero}, {Charlton}, \&  {Sembach} 2003]{gmcs03}
{Ganguly}, R., {Masiero}, J., {Charlton}, J.~C., \& {Sembach}, K.~R. 2003,  \apj, 598, 922

\bibitem[{Goodman} 2003]{goo03}
{Goodman}, J. 2003, \mnras, 339, 937

\bibitem[{Gunn}, {Siegmund}, {Mannery}, {Owen}, {Hull},  {Leger}, {Carey}, {Knapp}, {York}, {Boroski}, {Kent}, {Lupton}, {Rockosi},  {Evans}, {Waddell}, {Anderson}, {Annis}, {Barentine}, {Bartoszek}, {Bastian},  {Bracker}, {Brewington}, {Briegel}, {Brinkmann}, {Brown}, {Carr},  {Czarapata}, {Drennan}, {Dombeck}, {Federwitz}, {Gillespie}, {Gonzales},  {Hansen}, {Harvanek}, {Hayes}, {Jordan}, {Kinney}, {Klaene}, {Kleinman},  {Kron}, {Kresinski}, {Lee}, {Limmongkol}, {Lindenmeyer}, {Long}, {Loomis},  {McGehee}, {Mantsch}, {Neilsen}, {Neswold}, {Newman}, {Nitta}, {Peoples},  {Pier}, {Prieto}, {Prosapio}, {Rivetta}, {Schneider}, {Snedden}, \&  {Wang} 2006]{gun06}
{Gunn}, J., {Siegmund}, W., {Mannery}, E., {Owen}, R., {Hull}, C., {Leger}, R.,  {Carey}, L., {Knapp}, G., {et al.} 2006, \aj, 131, 2332

\bibitem[{Gunn}, {Carr}, {Rockosi}, {Sekiguchi}, {Berry},  {Elms}, {de Haas}, {Ivezi{\'c} }, {Knapp}, {Lupton}, {Pauls}, {Simcoe},  {Hirsch}, {Sanford}, {Wang}, {York}, {Harris}, {Annis}, {Bartozek},  {Boroski}, {Bakken}, {Haldeman}, {Kent}, {Holm}, {Holmgren}, {Petravick},  {Prosapio}, {Rechenmacher}, {Doi}, {Fukugita}, {Shimasaku}, {Okada}, {Hull},  {Siegmund}, {Mannery}, {Blouke}, {Heidtman}, {Schneider}, {Lucinio}, \&  {Brinkman} 1998]{gun98}
{Gunn}, J.~E., {Carr}, M., {Rockosi}, C., {Sekiguchi}, M., {Berry}, K., {Elms},  B., {de Haas}, E., {Ivezi{\'c} }, {\v Z}., {et al.} 1998, \aj,  116, 3040

\bibitem[{Hall}, {Anderson}, {Strauss}, {York},  {Richards}, {Fan}, {Knapp}, {Schneider}, {Vanden Berk}, \&  {Geballe} 2002]{sdss123}
{Hall}, P.~B., {Anderson}, S., {Strauss}, M., {York}, D., {Richards}, G.,  {Fan}, X., {Knapp}, G., {Schneider}, D., {et al.} 2002, \apjs, 141, 267

\bibitem[{Hall}, {Hutsem{\'e}kers}, {Anderson},  {Brinkmann}, {Fan}, {Schneider}, \& {York} 2003]{sb2}
{Hall}, P.~B., {Hutsem{\'e}kers}, D., {Anderson}, S.~F., {Brinkmann}, J.,  {Fan}, X., {Schneider}, D.~P., \& {York}, D.~G. 2003, \apj, 593, 189

\bibitem[{Hamann}, {Barlow}, {Cohen},  {Junkkarinen}, \& {Burbidge} 1997a]{hea97}
{Hamann}, F., {Barlow}, T., {Cohen}, R., {Junkkarinen}, V., \& {Burbidge}, E.  1997a, in Mass Ejection from Active Galactic Nuclei, ed. N.  Arav, I. Shlosman, \& R. Weymann (San Francisco: ASP), 19

\bibitem[{Hamann}, {Barlow}, \&  {Junkkarinen} 1997b]{hbj97}
{Hamann}, F., {Barlow}, T., \& {Junkkarinen}, V. 1997b, \apj, 478,  87

\bibitem[{Hamann}, {Barlow}, {Beaver}, {Burbidge},  {Cohen}, {Junkkarinen}, \& {Lyons} 1995]{hea95}
{Hamann}, F., {Barlow}, T.~A., {Beaver}, E.~A., {Burbidge}, E.~M., {Cohen},  R.~D., {Junkkarinen}, V., \& {Lyons}, R. 1995, \apj, 443, 606

\bibitem[{Hogg}, {Finkbeiner}, {Schlegel}, \&  {Gunn} 2001]{sdss82}
{Hogg}, D., {Finkbeiner}, D., {Schlegel}, D., \& {Gunn}, J. 2001, \aj, 122,  2129

\bibitem[{Horne} 1986]{hor86}
{Horne}, K. 1986, \pasp, 98, 609

\bibitem[{Ivezi{\'c}}, {Lupton}, {Schlegel},  {Boroski}, {Adelman-McCarthy}, {Yanny}, {Kent}, {Stoughton}, {Finkbeiner},  {Padmanabhan}, {Rockosi}, {Gunn}, {Knapp}, {Strauss}, {Richards},  {Eisenstein}, {Nicinski}, {Kleinman}, {Krzesinski}, {Newman}, {Snedden},  {Thakar}, {Szalay}, {Munn}, {Smith}, {Tucker}, \& {Lee} 2004]{ive04}
{Ivezi{\'c}}, {\v Z}., {Lupton}, R., {Schlegel}, D., {Boroski}, B.,  {Adelman-McCarthy}, J., {Yanny}, B., {Kent}, S., {Stoughton}, C., {et al.} 2004, AN, 325, 583

\bibitem[{Kendall} \& {Moran} 1963]{galp}
{Kendall}, M.~G. \& {Moran}, P. A.~P. 1963, {Geometrical Probability} (New  York: Hafner), 112

\bibitem[{Koratkar}, {Goad}, {O'Brien}, {Salamanca},  {Wanders}, {Axon}, {Crenshaw}, {Robinson}, {Korista}, {Rodriguez-Pascual},  {Horne}, {Blackwell}, {Carini}, {England}, {Perez}, {Pitts}, {Rawley},  {Reichert}, {Shrader}, \& {Wamsteker} 1996]{kea96}
{Koratkar}, A., {Goad}, M., {O'Brien}, P., {Salamanca}, I., {Wanders}, I.,  {Axon}, D., {Crenshaw}, D., {Robinson}, A., {et al.} 1996, \apj, 470, 378

\bibitem[{Laor}, {Barth}, {Ho}, \&  {Filippenko} 2006]{lbhf06}
{Laor}, A., {Barth}, A., {Ho}, L., \& {Filippenko}, A. 2006, \apj, 636, 83

\bibitem[{Leighly}, {Casebeer}, {Hamann}, \&  {Grupe} 2005]{lchg05}
{Leighly}, K.~M., {Casebeer}, D.~A., {Hamann}, F., \& {Grupe}, D. 2005, in  Bulletin of the American Astronomical Society, 1184

\bibitem[{Ma} 2002]{ma02}
{Ma}, F. 2002, \mnras, 335, L99

\bibitem[{Michalitsianos}, {Oliversen}, \&  {Nichols} 1996]{mon96}
{Michalitsianos}, A.~G., {Oliversen}, R.~J., \& {Nichols}, J. 1996, \apj, 461,  593

\bibitem[{Misawa}, {Eracleous}, {Charlton}, \&  {Tajitsu} 2005]{mect05}
{Misawa}, T., {Eracleous}, M., {Charlton}, J.~C., \& {Tajitsu}, A. 2005, \apj,  629, 115

\bibitem[{Murray}, {Chiang}, {Grossman}, \&  {Voit} 1995]{mur95}
{Murray}, N., {Chiang}, J., {Grossman}, S., \& {Voit}, G. 1995, \apj, 451, 498

\bibitem[{Owocki}, {Castor}, \& {Rybicki} 1988]{ocr88}
{Owocki}, S.~P., {Castor}, J.~I., \& {Rybicki}, G.~B. 1988, \apj, 335, 914

\bibitem[{Pereyra}, {Owocki}, {Hillier}, \&  {Turnshek} 2004]{poht04}
{Pereyra}, N.~A., {Owocki}, S.~P., {Hillier}, D.~J., \& {Turnshek}, D.~A. 2004,  \apj, 608, 454

\bibitem[{Peterson}, {Bentz}, {Desroches},  {Filippenko}, {Ho}, {Kaspi}, {Laor}, {Maoz}, {Moran}, {Pogge}, \&  {Quillen} 2006]{pee06}
{Peterson}, B., {Bentz}, M., {Desroches}, L.-B., {Filippenko}, A., {Ho}, L.,  {Kaspi}, S., {Laor}, A., {Maoz}, D., {et al.} 2006, \apj, 641, 638

\bibitem[{Peterson}, {Ferrarese}, {Gilbert}, {Kaspi},  {Malkan}, {Maoz}, {Merritt}, {Netzer}, {Onken}, {Pogge}, {Vestergaard}, \&  {Wandel} 2004]{pea04}
{Peterson}, B., {Ferrarese}, L., {Gilbert}, K., {Kaspi}, S., {Malkan}, M.,  {Maoz}, D., {Merritt}, D., {Netzer}, H., {et al.} 2004, \apj, 613, 682

\bibitem[{Peterson} 1997]{pet97}
{Peterson}, B.~M. 1997, Active Galactic Nuclei (Cambridge: Cambridge University  Press), 71--89

\bibitem[{Pier}, {Munn}, {Hindsley}, {Hennessy}, {Kent},  {Lupton}, \& {Ivezi{\' c}} 2003]{sdss153}
{Pier}, J.~R., {Munn}, J.~A., {Hindsley}, R.~B., {Hennessy}, G.~S., {Kent},  S.~M., {Lupton}, R.~H., \& {Ivezi{\' c}}, {\v Z}. 2003, \aj, 125, 1559

\bibitem[{Reichard}, {Richards}, {Schneider}, {Hall},  {Tolea}, {Krolik}, {Tsvetanov}, {Vanden Berk}, {York}, {Knapp}, {Gunn}, \&  {Brinkmann} 2003]{sdssbalcat}
{Reichard}, T., {Richards}, G., {Schneider}, D., {Hall}, P., {Tolea}, A.,  {Krolik}, J., {Tsvetanov}, Z., {Vanden Berk}, D., {et al.} 2003, \aj, 125, 1711

\bibitem[{Richards}, {Fan}, {Newberg}, {Strauss},  {Vanden Berk}, {Schneider}, {Yanny}, {Boucher}, \& {Burles} 2002]{sdssqtarget}
{Richards}, G., {Fan}, X., {Newberg}, H., {Strauss}, M., {Vanden Berk}, D.,  {Schneider}, D., {Yanny}, B., {Boucher}, A., {et al.} 2002, \aj, 123,  2945

\bibitem[Robbins 1944]{gplin}
Robbins, H.~E. 1944, Annals of Mathematical Statistics, 15, 70

\bibitem[{Rupke}, {Veilleux}, \& {Sanders} 2002]{rvs02}
{Rupke}, D.~S., {Veilleux}, S., \& {Sanders}, D.~B. 2002, \apj, 570, 588

\bibitem[{Rupke}, {Veilleux}, \& {Sanders} 2005]{rvs05}
---. 2005, \apjs, 160, 87

\bibitem[{Schneider}, {Hall}, {Richards}, {Vanden  Berk}, {Anderson}, {Fan}, {Jester}, {Stoughton}, {Strauss}, {SubbaRao},  {Brandt}, {Gunn}, {Yanny}, {Bahcall}, {Barentine}, {Blanton}, {Boroski},  {Brewington}, {Brinkmann}, {Brunner}, {Csabai}, {Doi}, {Eisenstein},  {Frieman}, {Fukugita}, {Gray}, {Harvanek}, {Heckman}, {Ivezi{\'c}}, {Kent},  {Kleinman}, {Knapp}, {Kron}, {Krzesinski}, {Long}, {Loveday}, {Lupton},  {Margon}, {Munn}, {Neilsen}, {Newberg}, {Newman}, {Nichol}, {Nitta}, {Pier},  {Rockosi}, {Saxe}, {Schlegel}, {Snedden}, {Szalay}, {Thakar}, {Uomoto},  {Voges}, \& {York} 2005]{dr3q}
{Schneider}, D.~P., {Hall}, P.~B., {Richards}, G.~T., {Vanden Berk}, D.~E.,  {Anderson}, S.~F., {Fan}, X., {Jester}, S., {Stoughton}, C., {et al.} 2005, \aj, 130, 367

\bibitem[{Schneider}, {Richards}, {Fan}, {Hall},  {Strauss}, {Vanden Berk}, {Gunn}, {Newberg}, \& {Reichard} 2002]{sdssedrq}
{Schneider}, D.~P., {Richards}, G.~T., {Fan}, X., {Hall}, P.~B., {Strauss},  M.~A., {Vanden Berk}, D.~E., {Gunn}, J.~E., {Newberg}, H.~J., {et al.} 2002, \aj, 123, 567

\bibitem[{Shakura} \& {Sunyaev} 1973]{ss73}
{Shakura}, N.~I. \& {Sunyaev}, R.~A. 1973, \aap, 24, 337

\bibitem[{Sirko} \& {Goodman} 2003]{sg03}
{Sirko}, E. \& {Goodman}, J. 2003, \mnras, 341, 501

\bibitem[{Sluse}, {Claeskens}, {Hutsem{\'e}kers}, \& {Surdej} 2007]{schs07}
{Sluse}, D., {Claeskens}, J.-F., {Hutsem{\'e}kers}, D., \& {Surdej}, J., \aap,
in press (astro-ph/0703030)

\bibitem[{Smith}, {Tucker}, {Kent}, {Richmond},  {Fukugita}, {Ichikawa}, {Ichikawa}, {Jorgensen}, {Uomoto}, {Gunn}, {Hamabe},  {Watanabe}, {Tolea}, {Henden}, {Annis}, {Pier}, {McKay}, {Brinkmann}, {Chen},  {Holtzman}, {Shimasaku}, \& {York} 2002]{sdss105}
{Smith}, J., {Tucker}, D., {Kent}, S., {Richmond}, M., {Fukugita}, M.,  {Ichikawa}, T., {Ichikawa}, S., {Jorgensen}, A., {et al.} 2002, \aj, 123, 2121

\bibitem[{Smith}, {Robinson}, {Young}, {Axon}, \&  {Corbett} 2005]{sea05}
{Smith}, J.~E., {Robinson}, A., {Young}, S., {Axon}, D.~J., \& {Corbett}, E.~A.  2005, \mnras, 359, 846

\bibitem[{Smith} \& {Penston} 1988]{sp88}
{Smith}, L.~J. \& {Penston}, M.~V. 1988, \mnras, 235, 551

\bibitem[{Srianand}, {Petitjean}, {Ledoux}, \&  {Hazard} 2002]{splh02}
{Srianand}, R., {Petitjean}, P., {Ledoux}, C., \& {Hazard}, C. 2002, \mnras,  336, 753

\bibitem[{Trump}, {Hall}, {Reichard}, {Richards},  {Schneider}, {Vanden Berk}, {Knapp}, {Anderson}, {Fan}, {Brinkman},  {Kleinman}, \& {Nitta} 2006]{trump06}
{Trump}, J., {Hall}, P., {Reichard}, T., {Richards}, G., {Schneider}, D.,  {Vanden Berk}, D., {Knapp}, G., {Anderson}, S., {et al.} 2006, \apjs, 165, 1

\bibitem[{Turnshek}, {Grillmair}, {Foltz}, \&  {Weymann} 1988]{tea88}
{Turnshek}, D.~A., {Grillmair}, C.~J., {Foltz}, C.~B., \& {Weymann}, R.~J.  1988, \apj, 325, 651

\bibitem[{Vanden Berk}, {Richards}, {Bauer},  {Strauss}, {Schneider}, {Heckman}, {York}, {Hall}, {Fan}, {Knapp},  {Anderson}, {Annis}, {Bahcall}, {Bernardi}, {Briggs}, {Brinkmann}, {Brunner},  {Burles}, {Carey}, {Castander}, {Connolly}, {Crocker}, {Csabai}, {Doi},  {Finkbeiner}, {Friedman}, {Frieman}, {Fukugita}, {Gunn}, {Hennessy},  {Ivezi{\' c}}, {Kent}, {Kunszt}, {Lamb}, {Leger}, {Long}, {Loveday},  {Lupton}, {Meiksin}, {Merelli}, {Munn}, {Newberg}, {Newcomb}, {Nichol},  {Owen}, {Pier}, {Pope}, {Rockosi}, {Schlegel}, {Siegmund}, {Smee}, {Snir},  {Stoughton}, {Stubbs}, {SubbaRao}, {Szalay}, {Szokoly}, {Tremonti}, {Uomoto},  {Waddell}, {Yanny}, \& {Zheng} 2001]{sdss73}
{Vanden Berk}, D.~E., {Richards}, G.~T., {Bauer}, A., {Strauss}, M.~A.,  {Schneider}, D.~P., {Heckman}, T.~M., {York}, D.~G., {Hall}, P.~B., {et al.} 2001, \aj, 122, 549

\bibitem[{Vestergaard}, {Wilkes}, \&  {Barthel} 2000]{vwb00}
{Vestergaard}, M., {Wilkes}, B.~J., \& {Barthel}, P.~D. 2000, \apjl, 538, L103

\bibitem[{Vilkoviskij} \& {Irwin} 2001]{vi01}
{Vilkoviskij}, E.~Y. \& {Irwin}, M.~J. 2001, \mnras, 321, 4

\bibitem[{Voit}, {Shull}, \& {Begelman} 1987]{vsb87}
{Voit}, G.~M., {Shull}, J.~M., \& {Begelman}, M.~C. 1987, \apj, 316, 573

\bibitem[{Voit}, {Weymann}, \& {Korista} 1993]{vwk93}
{Voit}, G.~M., {Weymann}, R.~J., \& {Korista}, K.~T. 1993, \apj, 413, 95

\bibitem[{York}, {Adelman}, {Anderson}, {Anderson},  {Annis}, {Bahcall}, {Bakken}, {Barkhouser}, {Bastian}, {Berman}, {Boroski},  {Bracker}, {Briegel}, {Briggs}, {Brinkmann}, {Brunner}, {Burles}, {Carey},  {Carr}, {Castander}, {Chen}, {Colestock}, {Connolly}, {Crocker}, {Csabai},  {Czarapata}, {Davis}, {Doi}, {Dombeck}, {Eisenstein}, {Ellman}, {Elms},  {Evans}, {Fan}, {Federwitz}, {Fiscelli}, {Friedman}, {Frieman}, {Fukugita},  {Gillespie}, {Gunn}, {Gurbani}, {de Haas}, {Haldeman}, {Harris}, {Hayes},  {Heckman}, {Hennessy}, {Hindsley}, {Holm}, {Holmgren}, {Huang}, {Hull},  {Husby}, {Ichikawa}, {Ichikawa}, {Ivezi{\'c}}, {Kent}, {Kim}, {Kinney},  {Klaene}, {Kleinman}, {Kleinman}, {Knapp}, {Korienek}, {Kron}, {Kunszt},  {Lamb}, {Lee}, {Leger}, {Limmongkol}, {Lindenmeyer}, {Long}, {Loomis},  {Loveday}, {Lucinio}, {Lupton}, {MacKinnon}, {Mannery}, {Mantsch}, {Margon},  {McGehee}, {McKay}, {Meiksin}, {Merelli}, {Monet}, {Munn}, {Narayanan},  {Nash}, {Neilsen}, {Neswold}, {Newberg}, {Nichol}, {Nicinski}, {Nonino},  {Okada}, {Okamura}, {Ostriker}, {Owen}, {Pauls}, {Peoples}, {Peterson},  {Petravick}, {Pier}, {Pope}, {Pordes}, {Prosapio}, {Rechenmacher}, {Quinn},  {Richards}, {Richmond}, {Rivetta}, {Rockosi}, {Ruthmansdorfer}, {Sandford},  {Schlegel}, {Schneider}, {Sekiguchi}, {Sergey}, {Shimasaku}, {Siegmund},  {Smee}, {Smith}, {Snedden}, {Stone}, {Stoughton}, {Strauss}, {Stubbs},  {SubbaRao}, {Szalay}, {Szapudi}, {Szokoly}, {Thakar}, {Tremonti}, {Tucker},  {Uomoto}, {Vanden Berk}, {Vogeley}, {Waddell}, {Wang}, {Watanabe},  {Weinberg}, {Yanny}, \& {Yasuda} 2000]{yor00}
{York}, D., {Adelman}, J., {Anderson}, J., {Anderson}, S., {Annis}, J.,  {Bahcall}, N., {Bakken}, J., {Barkhouser}, R., {et al.} 2000, \aj,  120, 1579

\end{thebibliography}
\end{document}